\documentclass[longauth]{aa} 

\usepackage[utf8]{inputenc}
\usepackage[varg]{txfonts}
\usepackage[breaklinks, colorlinks, citecolor=blue, linkcolor=blue, urlcolor=cyan]{hyperref}
\usepackage{color}
\usepackage{amsmath}
\usepackage{graphicx}
\usepackage[english]{babel}
\usepackage{float}
\usepackage{url}
\usepackage{subfigure}
\usepackage{longtable}
\usepackage{fancyhdr}
\usepackage[normalem]{ulem}
\usepackage{comment}
\usepackage{tablefootnote}
\usepackage{lscape}
\usepackage{siunitx}

\usepackage{natbib,twoopt}
\usepackage{array}     
\newcolumntype{C}{>{$}c<{$}}
\bibpunct{(}{)}{;}{a}{}{,}             
\makeatletter
  \newcommandtwoopt{\citeads}[3][][]{\href{http://adsabs.harvard.edu/abs/#3}%
    {\def\hyper@linkstart##1##2{}%
     \let\hyper@linkend\@empty\citealp[#1][#2]{#3}}}
  \newcommandtwoopt{\citepads}[3][][]{\href{http://adsabs.harvard.edu/abs/#3}%
    {\def\hyper@linkstart##1##2{}%
     \let\hyper@linkend\@empty\citep[#1][#2]{#3}}}
  \newcommandtwoopt{\citetads}[3][][]{\href{http://adsabs.harvard.edu/abs/#3}%
    {\def\hyper@linkstart##1##2{}%
     \let\hyper@linkend\@empty\citet[#1][#2]{#3}}}
  \newcommandtwoopt{\citeyearads}[3][][]%
    {\href{http://adsabs.harvard.edu/abs/#3}
    {\def\hyper@linkstart##1##2{}%
     \let\hyper@linkend\@empty\citeyear[#1][#2]{#3}}}
\makeatother

\makeatletter

\let\orgautoref\autoref
\renewcommand{\autoref}
        {\def\equationautorefname{Eq.}%
         \def\figureautorefname{Fig.}%
         \def\sectionautorefname{Sect.}%
         \def\subsectionautorefname{Sect.}%
         \def\subsubsectionautorefname{Sect.}%
         \orgautoref}

\makeatother
\begin{document}

\title{Improved characterization of the TOI-2141 system: a dense sub-Neptune with non-transiting inner and outer companions}

\titlerunning{Refining the TOI-2141 system}
\authorrunning{R. Luque et al.}

\author{
R.~Luque\inst{\ref{inst:1}}\,$^{\href{https://orcid.org/0000-0002-4671-2957}{\protect\includegraphics[height=0.19cm]{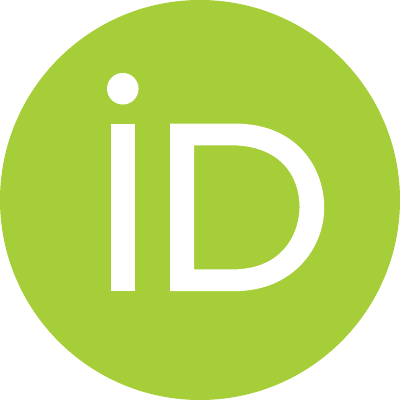}}}$\and 
K.~W.~F.~Lam\inst{\ref{inst:2}}\,$^{\href{https://orcid.org/0000-0002-9910-6088}{\protect\includegraphics[height=0.19cm]{figures/orcid.pdf}}}$\and 
J.~Cabrera\inst{\ref{inst:2}}\,$^{\href{https://orcid.org/0000-0001-6653-5487}{\protect\includegraphics[height=0.19cm]{figures/orcid.pdf}}}$\and 
A.~Bonfanti\inst{\ref{inst:3}}\,$^{\href{https://orcid.org/0000-0002-1916-5935}{\protect\includegraphics[height=0.19cm]{figures/orcid.pdf}}}$\and 
Y.~N.~E.~Eschen\inst{\ref{inst:4}}\,$^{\href{https://orcid.org/0009-0006-6397-2503}{\protect\includegraphics[height=0.19cm]{figures/orcid.pdf}}}$\and 
G.~Olofsson\inst{\ref{inst:5}}\,$^{\href{https://orcid.org/0000-0003-3747-7120}{\protect\includegraphics[height=0.19cm]{figures/orcid.pdf}}}$\and 
W.~Benz\inst{\ref{inst:6},\ref{inst:7}}\,$^{\href{https://orcid.org/0000-0001-7896-6479}{\protect\includegraphics[height=0.19cm]{figures/orcid.pdf}}}$\and 
N.~Billot\inst{\ref{inst:8}}\,$^{\href{https://orcid.org/0000-0003-3429-3836}{\protect\includegraphics[height=0.19cm]{figures/orcid.pdf}}}$\and 
A.~Brandeker\inst{\ref{inst:5}}\,$^{\href{https://orcid.org/0000-0002-7201-7536}{\protect\includegraphics[height=0.19cm]{figures/orcid.pdf}}}$\and 
A.~C.~M.~Correia\inst{\ref{inst:9}}\,$^{\href{https://orcid.org/0000-0002-8946-8579}{\protect\includegraphics[height=0.19cm]{figures/orcid.pdf}}}$\and 
L.~Fossati\inst{\ref{inst:3}}\,$^{\href{https://orcid.org/0000-0003-4426-9530}{\protect\includegraphics[height=0.19cm]{figures/orcid.pdf}}}$\and 
D.~Gandolfi\inst{\ref{inst:10}}\,$^{\href{https://orcid.org/0000-0001-8627-9628}{\protect\includegraphics[height=0.19cm]{figures/orcid.pdf}}}$\and 
H.~P.~Osborn\inst{\ref{inst:7},\ref{inst:11}}\,$^{\href{https://orcid.org/0000-0002-4047-4724}{\protect\includegraphics[height=0.19cm]{figures/orcid.pdf}}}$\and 
C.~Pezzotti\inst{\ref{inst:12}}\and 
S.~G.~Sousa\inst{\ref{inst:13}}\,$^{\href{https://orcid.org/0000-0001-9047-2965}{\protect\includegraphics[height=0.19cm]{figures/orcid.pdf}}}$\and 
T.~G.~Wilson\inst{\ref{inst:4}}\,$^{\href{https://orcid.org/0000-0001-8749-1962}{\protect\includegraphics[height=0.19cm]{figures/orcid.pdf}}}$\and 
S.~Wolf\inst{\ref{inst:14}}\and 
Y.~Alibert\inst{\ref{inst:7},\ref{inst:6}}\,$^{\href{https://orcid.org/0000-0002-4644-8818}{\protect\includegraphics[height=0.19cm]{figures/orcid.pdf}}}$\and 
R.~Alonso\inst{\ref{inst:15},\ref{inst:16}}\,$^{\href{https://orcid.org/0000-0001-8462-8126}{\protect\includegraphics[height=0.19cm]{figures/orcid.pdf}}}$\and 
J.~Asquier\inst{\ref{inst:17}}\and 
T.~Bárczy\inst{\ref{inst:18}}\,$^{\href{https://orcid.org/0000-0002-7822-4413}{\protect\includegraphics[height=0.19cm]{figures/orcid.pdf}}}$\and 
D.~Barrado\inst{\ref{inst:19}}\,$^{\href{https://orcid.org/0000-0002-5971-9242}{\protect\includegraphics[height=0.19cm]{figures/orcid.pdf}}}$\and 
S.~C.~C.~Barros\inst{\ref{inst:13},\ref{inst:20}}\,$^{\href{https://orcid.org/0000-0003-2434-3625}{\protect\includegraphics[height=0.19cm]{figures/orcid.pdf}}}$\and 
W.~Baumjohann\inst{\ref{inst:3}}\,$^{\href{https://orcid.org/0000-0001-6271-0110}{\protect\includegraphics[height=0.19cm]{figures/orcid.pdf}}}$\and 
F.~Biondi\inst{\ref{inst:21},\ref{inst:22}}\,$^{\href{https://orcid.org/0000-0002-1337-3653}{\protect\includegraphics[height=0.19cm]{figures/orcid.pdf}}}$\and 
L.~Borsato\inst{\ref{inst:22}}\,$^{\href{https://orcid.org/0000-0003-0066-9268}{\protect\includegraphics[height=0.19cm]{figures/orcid.pdf}}}$\and 
C.~Broeg\inst{\ref{inst:6},\ref{inst:7}}\,$^{\href{https://orcid.org/0000-0001-5132-2614}{\protect\includegraphics[height=0.19cm]{figures/orcid.pdf}}}$\and 
A.~Collier~Cameron\inst{\ref{inst:23}}\,$^{\href{https://orcid.org/0000-0002-8863-7828}{\protect\includegraphics[height=0.19cm]{figures/orcid.pdf}}}$\and 
Sz.~Csizmadia\inst{\ref{inst:2}}\,$^{\href{https://orcid.org/0000-0001-6803-9698}{\protect\includegraphics[height=0.19cm]{figures/orcid.pdf}}}$\and 
P.~E.~Cubillos\inst{\ref{inst:3},\ref{inst:24}}\and 
M.~B.~Davies\inst{\ref{inst:25}}\,$^{\href{https://orcid.org/0000-0001-6080-1190}{\protect\includegraphics[height=0.19cm]{figures/orcid.pdf}}}$\and 
M.~Deleuil\inst{\ref{inst:26}}\,$^{\href{https://orcid.org/0000-0001-6036-0225}{\protect\includegraphics[height=0.19cm]{figures/orcid.pdf}}}$\and 
A.~Deline\inst{\ref{inst:8}}\and 
O.~D.~S.~Demangeon\inst{\ref{inst:13},\ref{inst:20}}\,$^{\href{https://orcid.org/0000-0001-7918-0355}{\protect\includegraphics[height=0.19cm]{figures/orcid.pdf}}}$\and 
B.-O.~Demory\inst{\ref{inst:7},\ref{inst:27},\ref{inst:6}}\,$^{\href{https://orcid.org/0000-0002-9355-5165}{\protect\includegraphics[height=0.19cm]{figures/orcid.pdf}}}$\and 
A.~Derekas\inst{\ref{inst:28}}\and 
B.~Edwards\inst{\ref{inst:29}}\and 
J.~A.~Egger\inst{\ref{inst:6}}\,$^{\href{https://orcid.org/0000-0003-1628-4231}{\protect\includegraphics[height=0.19cm]{figures/orcid.pdf}}}$\and 
D.~Ehrenreich\inst{\ref{inst:8},\ref{inst:30}}\,$^{\href{https://orcid.org/0000-0001-9704-5405}{\protect\includegraphics[height=0.19cm]{figures/orcid.pdf}}}$\and 
A.~Erikson\inst{\ref{inst:2}}\and 
A.~Fortier\inst{\ref{inst:6},\ref{inst:7}}\,$^{\href{https://orcid.org/0000-0001-8450-3374}{\protect\includegraphics[height=0.19cm]{figures/orcid.pdf}}}$\and 
M.~Fridlund\inst{\ref{inst:31},\ref{inst:32}}\,$^{\href{https://orcid.org/0000-0002-0855-8426}{\protect\includegraphics[height=0.19cm]{figures/orcid.pdf}}}$\and 
K.~Gazeas\inst{\ref{inst:33}}\,$^{\href{https://orcid.org/0000-0002-8855-3923}{\protect\includegraphics[height=0.19cm]{figures/orcid.pdf}}}$\and 
M.~Gillon\inst{\ref{inst:34}}\,$^{\href{https://orcid.org/0000-0003-1462-7739}{\protect\includegraphics[height=0.19cm]{figures/orcid.pdf}}}$\and 
M.~Güdel\inst{\ref{inst:35}}\and 
M.~N.~Günther\inst{\ref{inst:17}}\,$^{\href{https://orcid.org/0000-0002-3164-9086}{\protect\includegraphics[height=0.19cm]{figures/orcid.pdf}}}$\and 
A.~Heitzmann\inst{\ref{inst:8}}\,$^{\href{https://orcid.org/0000-0002-8091-7526}{\protect\includegraphics[height=0.19cm]{figures/orcid.pdf}}}$\and 
Ch.~Helling\inst{\ref{inst:3},\ref{inst:36}}\and 
K.~G.~Isaak\inst{\ref{inst:17}}\,$^{\href{https://orcid.org/0000-0001-8585-1717}{\protect\includegraphics[height=0.19cm]{figures/orcid.pdf}}}$\and 
T.~Keller\inst{\ref{inst:6},\ref{inst:7}}\and 
L.~L.~Kiss\inst{\ref{inst:37},\ref{inst:38}}\and 
J.~Korth\inst{\ref{inst:8}}\,$^{\href{https://orcid.org/0000-0002-0076-6239}{\protect\includegraphics[height=0.19cm]{figures/orcid.pdf}}}$\and 
J.~Laskar\inst{\ref{inst:39}}\,$^{\href{https://orcid.org/0000-0003-2634-789X}{\protect\includegraphics[height=0.19cm]{figures/orcid.pdf}}}$\and 
A.~Lecavelier~des~Etangs\inst{\ref{inst:40}}\,$^{\href{https://orcid.org/0000-0002-5637-5253}{\protect\includegraphics[height=0.19cm]{figures/orcid.pdf}}}$\and 
A.~Leleu\inst{\ref{inst:8},\ref{inst:6}}\,$^{\href{https://orcid.org/0000-0003-2051-7974}{\protect\includegraphics[height=0.19cm]{figures/orcid.pdf}}}$\and 
M.~Lendl\inst{\ref{inst:8}}\,$^{\href{https://orcid.org/0000-0001-9699-1459}{\protect\includegraphics[height=0.19cm]{figures/orcid.pdf}}}$\and 
D.~Magrin\inst{\ref{inst:22}}\,$^{\href{https://orcid.org/0000-0003-0312-313X}{\protect\includegraphics[height=0.19cm]{figures/orcid.pdf}}}$\and 
G.~Mantovan\inst{\ref{inst:41},\ref{inst:22}}\and 
P.~F.~L.~Maxted\inst{\ref{inst:42}}\,$^{\href{https://orcid.org/0000-0003-3794-1317}{\protect\includegraphics[height=0.19cm]{figures/orcid.pdf}}}$\and 
B.~Merín\inst{\ref{inst:43}}\,$^{\href{https://orcid.org/0000-0002-8555-3012}{\protect\includegraphics[height=0.19cm]{figures/orcid.pdf}}}$\and 
C.~Mordasini\inst{\ref{inst:6},\ref{inst:7}}\and 
V.~Nascimbeni\inst{\ref{inst:22}}\,$^{\href{https://orcid.org/0000-0001-9770-1214}{\protect\includegraphics[height=0.19cm]{figures/orcid.pdf}}}$\and 
R.~Ottensamer\inst{\ref{inst:35}}\and 
I.~Pagano\inst{\ref{inst:44}}\,$^{\href{https://orcid.org/0000-0001-9573-4928}{\protect\includegraphics[height=0.19cm]{figures/orcid.pdf}}}$\and 
E.~Pallé\inst{\ref{inst:15},\ref{inst:16}}\,$^{\href{https://orcid.org/0000-0003-0987-1593}{\protect\includegraphics[height=0.19cm]{figures/orcid.pdf}}}$\and 
G.~Peter\inst{\ref{inst:2}}\,$^{\href{https://orcid.org/0000-0001-6101-2513}{\protect\includegraphics[height=0.19cm]{figures/orcid.pdf}}}$\and 
D.~Piazza\inst{\ref{inst:45}}\and 
G.~Piotto\inst{\ref{inst:22},\ref{inst:46}}\,$^{\href{https://orcid.org/0000-0002-9937-6387}{\protect\includegraphics[height=0.19cm]{figures/orcid.pdf}}}$\and 
D.~Pollacco\inst{\ref{inst:4}}\and 
D.~Queloz\inst{\ref{inst:11},\ref{inst:47}}\,$^{\href{https://orcid.org/0000-0002-3012-0316}{\protect\includegraphics[height=0.19cm]{figures/orcid.pdf}}}$\and 
R.~Ragazzoni\inst{\ref{inst:22},\ref{inst:46}}\,$^{\href{https://orcid.org/0000-0002-7697-5555}{\protect\includegraphics[height=0.19cm]{figures/orcid.pdf}}}$\and 
N.~Rando\inst{\ref{inst:17}}\and 
H.~Rauer\inst{\ref{inst:48},\ref{inst:49}}\,$^{\href{https://orcid.org/0000-0002-6510-1828}{\protect\includegraphics[height=0.19cm]{figures/orcid.pdf}}}$\and 
I.~Ribas\inst{\ref{inst:50},\ref{inst:51}}\,$^{\href{https://orcid.org/0000-0002-6689-0312}{\protect\includegraphics[height=0.19cm]{figures/orcid.pdf}}}$\and 
N.~C.~Santos\inst{\ref{inst:13},\ref{inst:20}}\,$^{\href{https://orcid.org/0000-0003-4422-2919}{\protect\includegraphics[height=0.19cm]{figures/orcid.pdf}}}$\and 
G.~Scandariato\inst{\ref{inst:44}}\,$^{\href{https://orcid.org/0000-0003-2029-0626}{\protect\includegraphics[height=0.19cm]{figures/orcid.pdf}}}$\and 
D.~Ségransan\inst{\ref{inst:8}}\,$^{\href{https://orcid.org/0000-0003-2355-8034}{\protect\includegraphics[height=0.19cm]{figures/orcid.pdf}}}$\and 
A.~E.~Simon\inst{\ref{inst:6},\ref{inst:7}}\,$^{\href{https://orcid.org/0000-0001-9773-2600}{\protect\includegraphics[height=0.19cm]{figures/orcid.pdf}}}$\and 
A.~M.~S.~Smith\inst{\ref{inst:2}}\,$^{\href{https://orcid.org/0000-0002-2386-4341}{\protect\includegraphics[height=0.19cm]{figures/orcid.pdf}}}$\and 
M.~Stalport\inst{\ref{inst:12},\ref{inst:34}}\and 
S.~Sulis\inst{\ref{inst:26}}\,$^{\href{https://orcid.org/0000-0001-8783-526X}{\protect\includegraphics[height=0.19cm]{figures/orcid.pdf}}}$\and 
Gy.~M.~Szabó\inst{\ref{inst:28},\ref{inst:52}}\,$^{\href{https://orcid.org/0000-0002-0606-7930}{\protect\includegraphics[height=0.19cm]{figures/orcid.pdf}}}$\and 
S.~Udry\inst{\ref{inst:8}}\,$^{\href{https://orcid.org/0000-0001-7576-6236}{\protect\includegraphics[height=0.19cm]{figures/orcid.pdf}}}$\and 
B.~Ulmer\inst{\ref{inst:2}}\and 
S.~Ulmer-Moll\inst{\ref{inst:53},\ref{inst:12}}\,$^{\href{https://orcid.org/0000-0003-2417-7006}{\protect\includegraphics[height=0.19cm]{figures/orcid.pdf}}}$\and 
V.~Van~Grootel\inst{\ref{inst:12}}\,$^{\href{https://orcid.org/0000-0003-2144-4316}{\protect\includegraphics[height=0.19cm]{figures/orcid.pdf}}}$\and 
J.~Venturini\inst{\ref{inst:8}}\,$^{\href{https://orcid.org/0000-0001-9527-2903}{\protect\includegraphics[height=0.19cm]{figures/orcid.pdf}}}$\and 
E.~Villaver\inst{\ref{inst:15},\ref{inst:16}}\and 
N.~A.~Walton\inst{\ref{inst:54}}\,$^{\href{https://orcid.org/0000-0003-3983-8778}{\protect\includegraphics[height=0.19cm]{figures/orcid.pdf}}}$\and 
D.~Wolter\inst{\ref{inst:2}}
}

\institute{
\label{inst:1} Instituto de Astrof{\'i}sica de Andaluc{\'i}a (IAA-CSIC), Glorieta de la Astronom{\'i}a s/n, 18008 Granada, Spain \and
\label{inst:2} Institute of Space Research, German Aerospace Center (DLR), Rutherfordstrasse 2, 12489 Berlin, Germany \and
\label{inst:3} Space Research Institute, Austrian Academy of Sciences, Schmiedlstrasse 6, A-8042 Graz, Austria \and
\label{inst:4} Department of Physics, University of Warwick, Gibbet Hill Road, Coventry CV4 7AL, United Kingdom \and
\label{inst:5} Department of Astronomy, Stockholm University, AlbaNova University Center, 10691 Stockholm, Sweden \and
\label{inst:6} Space Research and Planetary Sciences, Physics Institute, University of Bern, Gesellschaftsstrasse 6, 3012 Bern, Switzerland \and
\label{inst:7} Center for Space and Habitability, University of Bern, Gesellschaftsstrasse 6, 3012 Bern, Switzerland \and
\label{inst:8} Observatoire astronomique de l'Université de Genève, Chemin Pegasi 51, 1290 Versoix, Switzerland \and
\label{inst:9} CFisUC, Departamento de Física, Universidade de Coimbra, 3004-516 Coimbra, Portugal \and
\label{inst:10} Dipartimento di Fisica, Università degli Studi di Torino, via Pietro Giuria 1, I-10125, Torino, Italy \and
\label{inst:11} ETH Zurich, Department of Physics, Wolfgang-Pauli-Strasse 2, CH-8093 Zurich, Switzerland \and
\label{inst:12} Space sciences, Technologies and Astrophysics Research (STAR) Institute, Université de Liège, Allée du 6 Août 19C, 4000 Liège, Belgium \and
\label{inst:13} Instituto de Astrofisica e Ciencias do Espaco, Universidade do Porto, CAUP, Rua das Estrelas, 4150-762 Porto, Portugal \and
\label{inst:14} Space Research and Planetary 
Sciences, Physics Institute, University of Bern, Gesellschaftsstrasse 6, 3012 Bern, Switzerland \and
\label{inst:15} Instituto de Astrofísica de Canarias, Vía Láctea s/n, 38200 La Laguna, Tenerife, Spain \and
\label{inst:16} Departamento de Astrofísica, Universidad de La Laguna, Astrofísico Francisco Sanchez s/n, 38206 La Laguna, Tenerife, Spain \and
\label{inst:17} European Space Agency (ESA), European Space Research and Technology Centre (ESTEC), Keplerlaan 1, 2201 AZ Noordwijk, The Netherlands \and
\label{inst:18} Admatis, 5. Kandó Kálmán Street, 3534 Miskolc, Hungary \and
\label{inst:19} Depto. de Astrofísica, Centro de Astrobiología (CSIC-INTA), ESAC campus, 28692 Villanueva de la Cañada (Madrid), Spain \and
\label{inst:20} Departamento de Fisica e Astronomia, Faculdade de Ciencias, Universidade do Porto, Rua do Campo Alegre, 4169-007 Porto, Portugal \and
\label{inst:21} Max Planck Institute for Extraterrestrial Physics,Gießenbachstraße 1, 85748 Garching bei München \and
\label{inst:22} INAF, Osservatorio Astronomico di Padova, Vicolo dell'Osservatorio 5, 35122 Padova, Italy \and
\label{inst:23} Centre for Exoplanet Science, SUPA School of Physics and Astronomy, University of St Andrews, North Haugh, St Andrews KY16 9SS, UK \and
\label{inst:24} INAF, Osservatorio Astrofisico di Torino, Via Osservatorio, 20, I-10025 Pino Torinese To, Italy \and
\label{inst:25} Centre for Mathematical Sciences, Lund University, Box 118, 221 00 Lund, Sweden \and
\label{inst:26} Aix Marseille Univ, CNRS, CNES, LAM, 38 rue Frédéric Joliot-Curie, 13388 Marseille, France \and
\label{inst:27} ARTORG Center for Biomedical Engineering Research, University of Bern, Bern, Switzerland \and
\label{inst:28} ELTE Gothard Astrophysical Observatory, 9700 Szombathely, Szent Imre h. u. 112, Hungary \and
\label{inst:29} SRON Netherlands Institute for Space Research, Niels Bohrweg 4, 2333 CA Leiden, Netherlands \and
\label{inst:30} Centre Vie dans l’Univers, Faculté des sciences, Université de Genève, Quai Ernest-Ansermet 30, 1211 Genève 4, Switzerland \and
\label{inst:31} Leiden Observatory, University of Leiden, PO Box 9513, 2300 RA Leiden, The Netherlands \and
\label{inst:32} Department of Space, Earth and Environment, Chalmers University of Technology, Onsala Space Observatory, 439 92 Onsala, Sweden \and
\label{inst:33} National and Kapodistrian University of Athens, Department of Physics, University Campus, Zografos GR-157 84, Athens, Greece \and
\label{inst:34} Astrobiology Research Unit, Université de Liège, Allée du 6 Août 19C, B-4000 Liège, Belgium \and
\label{inst:35} Department of Astrophysics, University of Vienna, Türkenschanzstrasse 17, 1180 Vienna, Austria \and
\label{inst:36} Institute for Theoretical Physics and Computational Physics, Graz University of Technology, Petersgasse 16, 8010 Graz, Austria \and
\label{inst:37} Konkoly Observatory, Research Centre for Astronomy and Earth Sciences, 1121 Budapest, Konkoly Thege Miklós út 15-17, Hungary \and
\label{inst:38} ELTE E\"otv\"os Lor\'and University, Institute of Physics, P\'azm\'any P\'eter s\'et\'any 1/A, 1117 Budapest, Hungary \and
\label{inst:39} IMCCE, UMR8028 CNRS, Observatoire de Paris, PSL Univ., Sorbonne Univ., 77 av. Denfert-Rochereau, 75014 Paris, France \and
\label{inst:40} Institut d'astrophysique de Paris, UMR7095 CNRS, Université Pierre \& Marie Curie, 98bis blvd. Arago, 75014 Paris, France \and
\label{inst:41} Centro di Ateneo di Studi e Attività Spaziali G. Colombo -- Università degli Studi di Padova, Via Venezia 15, IT-35131, Padova, Italy \and
\label{inst:42} Astrophysics Group, Lennard Jones Building, Keele University, Staffordshire, ST5 5BG, United Kingdom \and
\label{inst:43} European Space Agency, ESA - European Space Astronomy Centre, Camino Bajo del Castillo s/n, 28692 Villanueva de la Cañada, Madrid, Spain \and
\label{inst:44} INAF, Osservatorio Astrofisico di Catania, Via S. Sofia 78, 95123 Catania, Italy \and
\label{inst:45} Weltraumforschung und Planetologie, Physikalisches Institut, University of Bern, Gesellschaftsstrasse 6, 3012 Bern, Switzerland \and
\label{inst:46} Dipartimento di Fisica e Astronomia "Galileo Galilei", Università degli Studi di Padova, Vicolo dell'Osservatorio 3, 35122 Padova, Italy \and
\label{inst:47} Cavendish Laboratory, JJ Thomson Avenue, Cambridge CB3 0HE, UK \and
\label{inst:48} German Aerospace Center (DLR), Markgrafenstrasse 37, 10117 Berlin, Germany \and
\label{inst:49} Institut fuer Geologische Wissenschaften, Freie Universitaet Berlin, Malteserstrasse 74-100,12249 Berlin, Germany \and
\label{inst:50} Institut de Ciencies de l'Espai (ICE, CSIC), Campus UAB, Can Magrans s/n, 08193 Bellaterra, Spain \and
\label{inst:51} Institut d'Estudis Espacials de Catalunya (IEEC), 08860 Castelldefels (Barcelona), Spain \and
\label{inst:52} HUN-REN-ELTE Exoplanet Research Group, Szent Imre h. u. 112., Szombathely, H-9700, Hungary \and
\label{inst:53} Leiden Observatory, University of Leiden, Einsteinweg 55, 2333 CA Leiden, The Netherlands \and
\label{inst:54} Institute of Astronomy, University of Cambridge, Madingley Road, Cambridge, CB3 0HA, United Kingdom
}

\date{Received 21 July 2025 / Accepted 29 August 2025}

\abstract
{}
{We aim to refine the fundamental parameters of the TOI-2141 planetary system, which includes a transiting sub-Neptune orbiting a Sun-like star in a relatively long orbit of 18.26\,days, by combining new photometric and spectroscopic observations.}
{We analyze new space-based photometry from TESS and CHEOPS as well as 61 radial velocity measurements from HARPS-N. We perform individual and joint photometric and RV analyses using several modeling tools within a Bayesian model comparison framework.}
{We refine the radius and mass of the transiting planet TOI-2141~b to $3.15\pm0.04\,R_\oplus$ and $20.1\pm1.6\,M_\oplus$, respectively, five and two times more precise than the previously reported values. Our radial velocity analysis reveals two additional non-transiting companions with orbital periods of 5.46 and 60.45 days. Despite the innermost planet's high geometric transit probability, we find no evidence for transits in the photometric data. }
{The bulk properties of TOI-2141~b suggest a significant volatile envelope atop an Earth-like core, with modeling indicating a hydrogen-rich atmosphere that may have experienced mild photoevaporation over the system's history. Planets b and c must exhibit a modest mutual inclination of at least $\sim 2.4^\circ$.}


\maketitle

\nolinenumbers
\section{Introduction} \label{sec:intro}


This work revisits the TOI-2141 planetary system, discovered and characterized by \citet{Martioli2023A&A...680A..84M} using space-based photometry from TESS \citep{TESS} and radial velocities (RVs) from the SOPHIE high-resolution spectrograph mounted on the 1.93-m telescope at the Observatoire de Haute-Provence \citep{Perruchot2008,Bouchy2013}. The star TOI-2141 is a relatively bright ($V=9.46\,\mathrm{mag}$), solar analog ($T_{\rm eff} \sim 5650\,\mathrm{K}$) hosting a transiting sub-Neptune ($R_p \sim 3\,R_\oplus$, $M_p \sim 24\,M_\oplus$) with an orbital period of 18.26\,days. The authors argue that TOI-2141~b may have a ``likely dense rocky core and a possible thick, water-rich envelope'', but several internal structure models could explain its measured bulk density \citep[as pointed out by, e.g.,][]{RogersSeager2010, 2014ApJ...792....1L, Zeng2016ApJ...819..127Z, NixonMadhusudhan2021, Luo2024}. Here, we add unpublished TESS observations from June 2024, four transits of TOI-2141~b with CHEOPS\footnote{Observed as part of the Guaranteed Time Observervation (GTO) programmes CH\_PR100024 and CH\_PR140073. The raw and detrended photometric time-series data are available in electronic form at the CDS via anonymous ftp to \texttt{cdsarc.u-strasbg.fr} (130.79.128.5) or via \url{http://cdsweb.u-strasbg.fr/cgi-bin/qcat?J/A+A/}} \citep{Benz2021}, and 61 RVs from HARPS-N \citep{HARPSN} to refine the properties of the system, including its architecture.

\section{Observations} \label{sec:data}

\subsection{TESS photometry} \label{subsec:tess}

\begin{figure*}[ht!]
    \centering
    \includegraphics[width=\hsize]{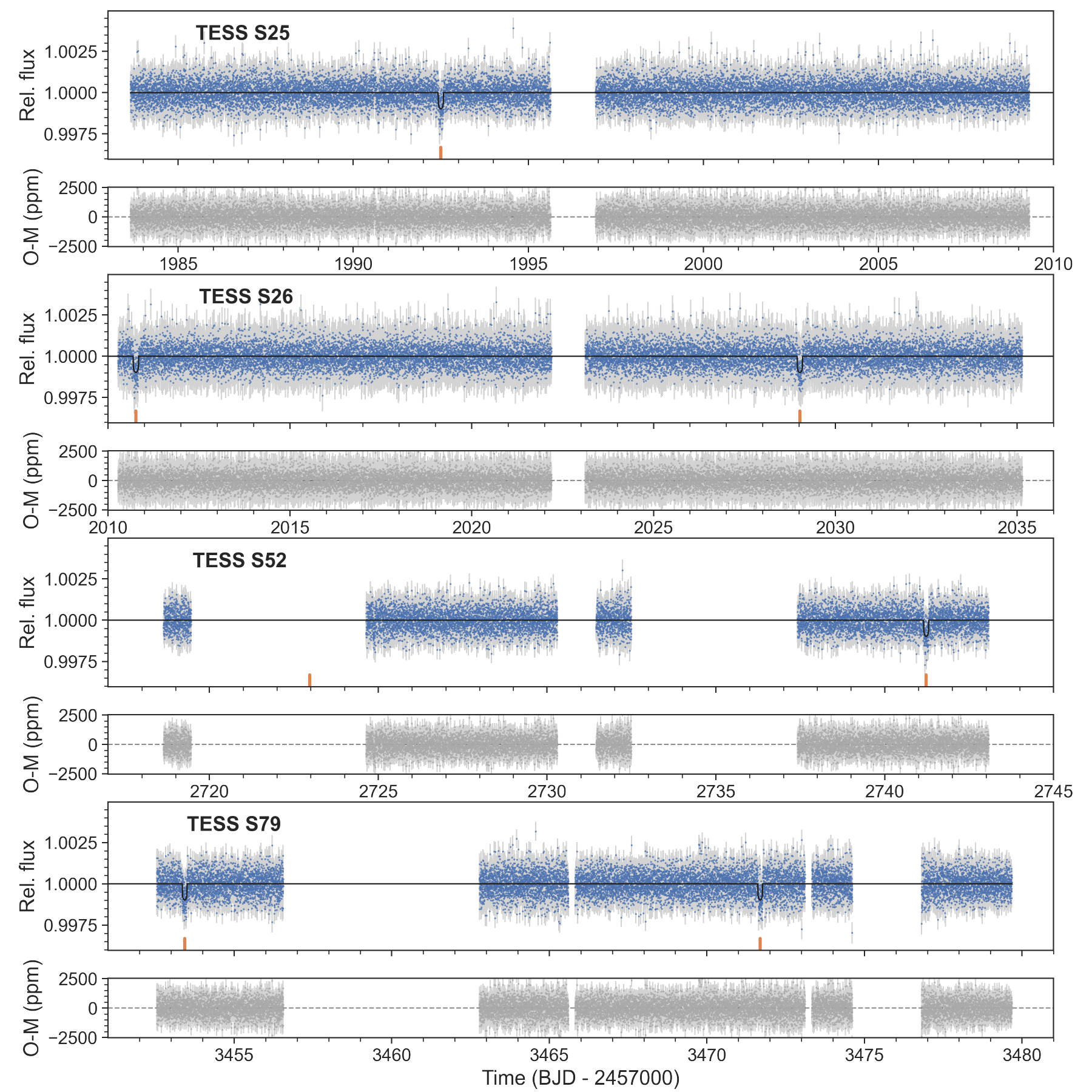}
    \caption{PDC-corrected photometry of TOI-2141 from TESS. Gray points are the original 2-min cadence TESS data, while blue shows 15-min binned photometric data. Transits of the planet TOI-2141~b are marked in orange. The black line shows the best-fit model to the data described in Sect.~\ref{subsec:joint_fit}.}
    \label{fig:tess_lc}
\end{figure*}

The TESS mission observed TOI-2141 with a cadence of 2-min in Sectors 25, 26, 52, and 79. Data from Sectors 25, 26, and 52 were described and analyzed in \citet{Martioli2023A&A...680A..84M}, where the transiting planet TOI-2141~b was identified and confirmed. In this work, we add to our analyses the photometry from Sector 79 gathered in June 2024 and combine it with the rest of the photometry. We downloaded the TESS data products from the Mikulski Archive for Space Telescopes\footnote{\url{https://mast.stci.edu}} and use the Presearch Data Conditioning (PDC) flux time series \citep{Smith2012PASP..124.1000S,Stumpe2012PASP..124..985S,Stumpe2014PASP..126..100S} computed by the TESS Science Processing Operations Center (SPOC) pipeline \citep{SPOC}. Figure~\ref{fig:tess_lc} shows the complete TESS dataset used in this work.

\subsection{CHEOPS photometry} \label{subsec:cheops}

\begin{table*}[t]
\caption{CHEOPS observing log.} \label{tab:cheops_dat}
\footnotesize{
\centering
\begin{tabular}{ccccccc}
\hline
\hline
\noalign{\smallskip}
Visit & Start (UTC) & Length [hr] & Archive filekey & No. Frames & RMS (ppm) & Efficiency (\%)\\
\noalign{\smallskip}
\hline
\noalign{\smallskip}
1 & 2023-05-05T03:38:19 & 13.29 & CH\_PR100024\_TG016301\_V0300 & 506 & 426.1 & 63.4 \\
2 & 2023-06-10T15:58:19 & 13.37 & CH\_PR100024\_TG016302\_V0300 & 639 & 300.0 & 79.5 \\
3 & 2023-06-28T22:15:19 & 13.34 & CH\_PR100024\_TG016303\_V0300 & 555 & 307.7 & 69.3  \\
4 & 2024-06-28T05:01:58 & 11.04 & CH\_PR140073\_TG002301\_V0300 & 517 & 315.0 & 77.9 \\
5 & 2024-07-16T10:37:59 & 11.84 & CH\_PR140073\_TG002302\_V0300 & 411 & 283.7 & 57.8 \\

\noalign{\smallskip}
\hline
\end{tabular}
}
\end{table*}

The CHaracterising ExOPlanets Satellite (CHEOPS) is a European Space Agency small-class mission dedicated to studying bright exoplanet host stars using high-precision photometric observations \citep{CHEOPS,Benz2021,Fortier2024}. CHEOPS observed TOI-2141 between 05 May 2023 and 16 July 2024. The observations were taken under the CHEOPS Guaranteed Time Observation (GTO) programmes PR100024 (\texttt{MR.Improve}) and PR140073 (\texttt{Gas Content of Low Mass Planets}). Five visits were observed with an exposure cadence of 60\,s. The data were extracted with the CHEOPS DRP v14 pipeline \citep{hoyer2020} which performed instrument calibrations, as well as cosmic rays, background and smearing corrections. Aperture sizes with radii between 15 and 40 pixels were extracted. We opted to use light curves extracted using aperture radii of either 22 or 23 pixels for further analyses, as these have the lowest RMS values.

Figure~\ref{fig:cheops_raw} shows the original CHEOPS data and the detrended light curves used in our subsequent analyses. Table~\ref{tab:cheops_dat} presents an observing log summarizing the properties of each visit.

\subsection{HARPS-N spectroscopy} \label{subsec:harps-n}

\begin{figure}
    \centering
    \includegraphics[width=\hsize]{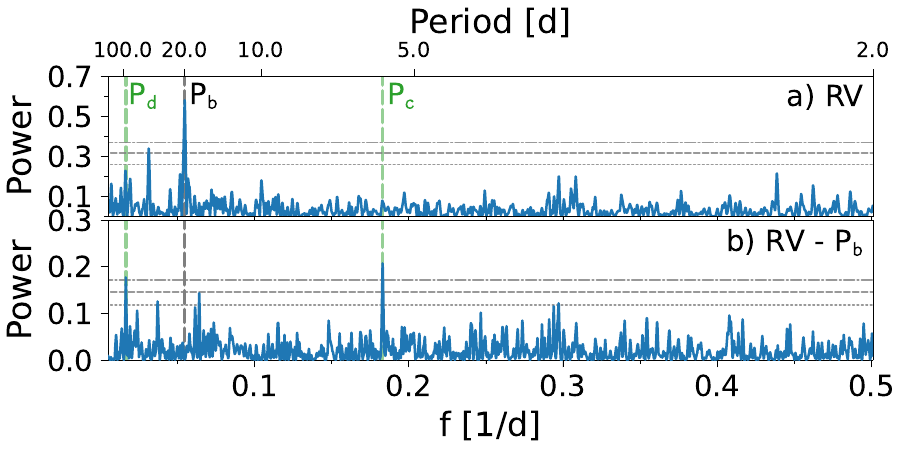}
    \caption{GLS periodograms of the SOPHIE and HARPS-N RV data. For each panel, the horizontal lines show the theoretical 10\% (short-dashed line), 1\% (long-dashed line), and 0.1\% (dot-dashed line) false alarm probability levels. The vertical dashed lines mark the orbital frequencies of the transiting planet ($f_\mathrm{b}=0.0547\,{\rm d}^{-1}$) and of the signals detected in the RVs at 5.46 and 60\,d. \emph{Panel a}: Combined RV data corrected for an instrumental offset. \emph{Panel b}: RV residuals following the subtraction of the signal of the transiting planet TOI-2141~b. }
    \label{fig:gls_rvs}
\end{figure}

We collected 61 RVs of TOI-2141 with the HARPS-N spectrograph mounted at the 3.6-m Telescopio Nazionale Galileo in La Palma between October 2022 and September 2024. The data was collected as part of the International Time Programme ``Completing CHEOPS characterization of long-period low-mass planets'' (PI: Luque; program IDs ITP22\_2 and ITP23\_10) granted by the International Scientific Committee of the Observatorios de Canarias of the IAC. Radial velocities and additional spectral indicators were derived using YABI \citep{Borsa2015A&A...578A..64B} and \texttt{serval} \cite{SERVAL}. YABI is an online implementation of the classical HARPS-N DRS pipeline at the INAF Trieste
Observatory\footnote{\url{https://www.ia2.inaf.it/}} and computes cross-correlation functions (CCFs) for a given template mask following \citet{2014SPIE.9147E..8CC}. On the other hand, \texttt{serval} follows a template-matching technique \citep{TERRA} to compute radial velocities from a high signal-to-noise (S/N) template built from the stellar spectra themselves. The YABI-derived radial velocities have a mean internal precision of $0.98\,\mathrm{m\,s^{-1}}$ and show a standard deviation of $4.7\,\mathrm{m\,s^{-1}}$. The \texttt{serval} radial velocities have a mean internal precision of $1.3\,\mathrm{m\,s^{-1}}$ and show the same standard deviation. Due to their superior precision, we use the YABI RVs in our subsequent analyses. 

Figure~\ref{fig:rvs} shows the HARPS-N RVs together with the SOPHIE RVs presented in \citet{Martioli2023A&A...680A..84M} used in this work. A search for periodic signals in the combined dataset (subtracting for an instrumental offset) using a Generalized Lomb-Scargle periodogram \citep{GLS} shows a prominent peak (false alarm probability below 10$^{-6}$) at the period of the transiting planet TOI-2141~b ($P_{\rm b}=18.26\,\mathrm{days}$, $f_{\rm b}=0.0547\,\mathrm{d^{-1}}$). Surprisingly, a periodogram on the residuals of a RV-only model as in \citet{Martioli2023A&A...680A..84M} that includes a circular orbit at the period of the transiting planet shows two peaks with false alarm probabilities (FAPs) below 0.1\% at 5.46 and 60\,days. We analyze these two signals in detail in Sect.~\ref{subsec:rv_fit}. 

A GLS periodogram of all the available activity indicators is shown in Fig.~\ref{fig:gls_activity} of the Appendix. The GLS periodograms do not find any significant peaks (FAP below 10\%) in any of the indicators. For H$\alpha$, a slight peak approaching FAP$\sim$10\% can be found at 40\,days, approximately double the inferred rotational period of TOI-2141 of $21\pm5$\,d \citep{Martioli2023A&A...680A..84M} based on empirical activity-rotation relations \citep{MamajekHillenbrand2008}. This peak is also seen in SOPHIE's H$\alpha$ index \citep[][their Fig. C.6]{Martioli2023A&A...680A..84M}, but again below the significance threshold. Compatible with the reported rotational period of approximately 20\,days, there is only a non-significant peak (FAP$>$10\%) in the full-width at half-maximum of the CCF of the HARPS-N data at 17\,d also seen in the bisector span of the SOPHIE CCFs \citep{Martioli2023A&A...680A..84M}. The lack of long-term continuous photometric monitoring of the star prevents us from refining the stellar rotation period using an independent technique.

\section{Analysis and Results} \label{sec:analyses}

\subsection{Stellar characterization} \label{subsec:stellar_properties}
\begin{table}
\centering
\small
\caption{Stellar parameters of TOI-2141.} \label{tab:star}
\begin{tabular}{lcr}
\hline\hline
\noalign{\smallskip}
Parameter                               & Value                 & Reference \\ 
\hline
\noalign{\smallskip}
\multicolumn{3}{c}{\it Name and identifiers}\\
\noalign{\smallskip}
Name                            & BD+18 3330                & BD    \\
TIC                             & 287256467                 & TIC   \\  
TOI                             & TOI-2141                  & {\citet{Martioli2023A&A...680A..84M}}      \\
\noalign{\smallskip}
\multicolumn{3}{c}{\it Coordinates and spectral type}\\
\noalign{\smallskip}
$\alpha$                                & 17:15:02.906      & Gaia EDR3     \\
$\delta$                                & +18:20:26.76      & Gaia EDR3     \\
Epoch (ICRS)                            & J2000             & Gaia EDR3     \\
\noalign{\smallskip}
\multicolumn{3}{c}{\it Magnitudes}\\
\noalign{\smallskip}
$V$ [mag]                               &  $9.46\pm0.003$       & UCAC4       \\
$G$ [mag]                               &  $9.34408\pm0.00021$  & Gaia EDR3   \\
$J$ [mag]                               &  $8.27\pm0.02$        & 2MASS       \\
$K_s$ [mag]                             &  $7.87\pm0.02$        & 2MASS       \\
\noalign{\smallskip}
\multicolumn{3}{c}{\it Parallax and kinematics}\\
\noalign{\smallskip}
$\pi$ [mas]                             & $12.957\pm0.015$      & Gaia EDR3             \\
$d$ [pc]                                & $77.7\pm0.2$          & Gaia EDR3             \\
$\mu_{\alpha}\cos\delta$ [$\mathrm{mas\,yr^{-1}}$]  & $52.229 \pm 0.013$    & Gaia EDR3          \\
$\mu_{\delta}$ [$\mathrm{mas\,yr^{-1}}$]            & $-98.286 \pm 0.015$   & Gaia EDR3          \\
$U$ [$\mathrm{km\,s^{-1}}]$             &  $15.56\pm0.30$       & This work      \\
$V$ [$\mathrm{km\,s^{-1}}]$             &  $-21.60\pm0.25$       & This work      \\
$W$ [$\mathrm{km\,s^{-1}}]$             &  $-36.53\pm0.22$       & This work      \\
\noalign{\smallskip}
\multicolumn{3}{c}{\it Photospheric parameters}\\
\noalign{\smallskip}
$T_{\mathrm{eff}}$ [K]                      & $5635 \pm 61$         & This work   \\
$\log (g~[\mathrm{cm\,s}^{-2}])$                                     & $4.40 \pm 0.03$       & This work   \\
{[Fe/H]}                                    & $-0.14 \pm 0.04$      & This work   \\
$v_{\mathrm{micro}}$ [$\mathrm{km\,s^{-1}}$]    & $0.92 \pm 0.01$   & This work             \\
\noalign{\smallskip}
\multicolumn{3}{c}{\it Physical parameters}\\
\noalign{\smallskip}
$M$ [$M_{\odot}$]                       & $0.896_{-0.051}^{+0.059}$         & This work       \\
$R$ [$R_{\odot}$]                       & $0.950 \pm 0.007$         & This work       \\
Age [Gyr]                               & $9 \pm 4$                 & This work       \\
\noalign{\smallskip}
\hline
\end{tabular}
\tablebib{
    BD: \citet{BD_Catalogue};
    TIC: \citet{Stassun2018AJ....156..102S};
    Gaia EDR3: \citet{GaiaEDR3};
    UCAC4: \citet{UCAC4};
    2MASS: \citet{2MASS}.
}
\end{table}

The stellar spectroscopic parameters ($T_{\mathrm{eff}}$, $\log g$, microturbulence, [Fe/H]) were derived using the ARES+MOOG methodology \citep[][]{Sousa-21, Sousa-14, Santos-13}. The equivalent widths (EW) were automatically measured using the ARES code\footnote{The last version, ARES v2, can be downloaded at \url{https://github.com/sousasag/ARES}} \citep{Sousa-07, Sousa-15}. In this spectral analysis, we used a combined HARPS-N spectrum from individual exposures of TOI-2141, which was then used to measure the EWs for a list of lines presented in \citet[][]{Sousa-08}. We analyze only the HARPS-N combined spectrum as it is of higher resolution (115,000 vs. 75,000) and S/N (630 vs. 507 on average) than the SOPHIE one. Besides, the SOPHIE spectrum was analyzed in detail in \citet{Martioli2023A&A...680A..84M}, for which we find below very consistent stellar parameters to ours. The best set of spectroscopic parameters was found by using a minimization process to find the ionization and excitation equilibrium. This process makes use of a grid of Kurucz model atmospheres \citep{Kurucz-93} and the latest version of the radiative transfer code MOOG \citep{Sneden-73}. We also derived a more accurate trigonometric surface gravity using recent Gaia EDR3 data following the same procedure as described in \citet[][]{Sousa-21}, which provided a consistent value compared to the spectroscopic surface gravity.


The stellar radius of TOI-2141 was computed using a MCMC modified infrared flux method \citep[IRFM --][]{Blackwell1977,Schanche2020}. By constructing spectral energy distributions (SED) from two stellar atmospheric models catalogues \citep{Kurucz-93,Castelli2003} that were constrained by our spectroscopically derived stellar parameters, we determined broadband synthetic photometry. Via comparison with observed fluxes in the following bandpasses:  2MASS $J$, $H$, and $K$, WISE $W1$ and $W2$, and Gaia $G$, $G_\mathrm{BP}$, and $G_\mathrm{RP}$ \citep{2MASS,Wright2010,GaiaEDR3}, we computed the stellar bolometric flux that we converted into the effective temperature and angular diameter of TOI-2141. Lastly, we derived the stellar radius by combining the angular diameter with the offset-corrected Gaia parallax \citep{Lindegren2021}. To account for stellar atmosphere model uncertainties, we conducted a Bayesian Model Averaging of the stellar radius posteriors produced when using each catalogue. The weighted average value ($0.950 \pm 0.007\,R_\odot$) is reported in Table 2.

We used $T_{\mathrm{eff}}$, [Fe/H], and $R_{\star}$ to derive the stellar mass $M_{\star}$ and age $t_{\star}$ accounting for two different stellar evolutionary models. In detail, we derive a first set of mass and age estimates thanks to the isochrone placement routine \citep{bonfanti2015,bonfanti2016} that interpolates the input parameters within pre-computed grids of PARSEC\footnote{\textsl{PA}dova and T\textsl{R}ieste \textsl{S}tellar \textsl{E}volutionary \textsl{C}ode: \url{https://stev.oapd.inaf.it/cgi-bin/cmd}} v1.2S \citep{marigo2017} tracks and isochrones. A second mass and age set, instead, was computed by the CLES \citep[Code Liègeois d'Évolution Stellaire;][]{scuflaire2008} code that builds the best-fit evolutionary track based on the input parameters following a Levenberg-Marquardt optimisation scheme \citep{salmon2021}. After carefully checking the consistency of the two respective pairs of outcomes via the $\chi^2$-based criterion outlined in \citet{bonfanti2021}, we finally merged (i.e. we summed) the respective output distributions and we obtained $M_{\star}=0.896_{-0.051}^{+0.059}\,M_{\odot}$ and $t_{\star}=9\pm4$ Gyr \citep[see][for further details]{bonfanti2021}.

Furthermore, we computed the right-handed, heliocentric Galactic space velocities \citep{Johnson1987}, $U$, $V$, and $W$, using the Gaia EDR3 coordinates, proper motions, offset-corrected parallax \citep{Lindegren2021}, and RV \citep{GaiaEDR3}. These are reported in Table~\ref{tab:star} and strongly indicate that the star is a member of the thin disk population. 

\subsection{CHEOPS analysis} \label{subsec:CHEOPS_fit}
We used the publicly available \texttt{pycheops} package \citep{maxted2022} to model systematic trends and fit transit light curves simultaneously. A 5-sigma clipping was first applied to the raw light curves to remove any outliers. Each visit was then detrended separately, where the set of detrending basis vectors used in respective detrending models were chosen based on the Bayes factor. A ‘glint’ effect can be introduced into the light curve due to internal reflection of bright objects (e.g. moonlight) within 24$^{\circ}$ from the observed target \citep{maxted2022}. We applied the \texttt{glint function} to remove glint effects using a spline fit with 8 to 11 knots. A Gaussian white noise component was also included in the model. \texttt{pycheops} implements the \texttt{LMFIT} package \citep{Newville2014} to perform nonlinear least-square minimisation and the \texttt{emcee} \citep{dfm2013} was used for Bayesian analysis and posterior sampling. The final detrended light curves are shown in Figure~\ref{fig:cheops_raw} and were used in subsequent joint fits.

\subsection{Radial velocity analysis} \label{subsec:rv_fit}

\begin{table}
    \centering
    \caption{Model comparison of RV-only fits with \texttt{juliet}. }  \label{tab:model_comparison}
    \begin{tabular}{lcc}
        \hline
        \hline
        \noalign{\smallskip}
        Model & $\ln Z$ & $K_{\rm b}$ (m\,s$^{-1}$)  \\
        \noalign{\smallskip}
        \hline
        \noalign{\smallskip}
1p          & $-475.0$   & $5.4\pm0.5$      \\
1p\_ecc     & $-475.0$   & $5.4\pm0.5$    \\
1p + GP     & $-473.3$   & $5.3\pm0.4$    \\[0.2cm]
2p  (5-d)   & $-470.8$   & $5.6\pm0.4$    \\
2p (60-d)   & $-470.9$   & $5.2\pm0.4$    \\
2p + GP     & $-463.9$   & $5.3\pm0.3$    \\[0.2cm]
{\bf 3p}    & $\mathbf{-459.3}$   & $\mathbf{5.3\pm0.3}$     \\
3p + GP     & $-465.5$   & $5.4\pm0.3$     \\
        \noalign{\smallskip}
        \hline
    \end{tabular}
\tablefoot{
    The GP refers to a Gaussian Process with an exponential sine-squared kernel from \texttt{george}. The model used for the final joint fit is marked in boldface.
}
\end{table}

We used \texttt{juliet} \citep{juliet}, a python library built on many publicly available tools for the modeling of transits \citep[\texttt{batman},][]{batman}, RVs \citep[\texttt{radvel},][]{radvel}, and Gaussian Processes (\texttt{george}, \citealt{Ambikasaran2015ITPAM..38..252A}; \texttt{celerite}, \citealt{celerite}) to model the RV data first. We use the nested sampler \texttt{dynesty} \citep{dynesty} to explore the parameter space of a given prior volume and to compute efficiently the Bayesian model log evidence ($\ln Z$). Thanks to this, we can compare models with different numbers of parameters accounting for the model complexity and the number of degrees of freedom with a sound statistical methodology. In our analysis, we consider that a model (M2) is greatly favored over another (M1) if $\Delta \ln Z = \ln Z_{\rm M2} - \ln Z_{M1} > 5$ \citep{2008ConPh..49...71T}. If $\Delta \ln Z \in [1,2.5)$, we consider that M2 is weakly favored over M1 so the simpler model with less degrees of freedom would be chosen. For intermediate cases, we consider that M2 is moderately favored over M1. 

We carry out a Bayesian model comparison analysis to assess the statistical significance of the RV signals present in the SOPHIE and HARPS-N data seen in Fig.~\ref{fig:gls_rvs}. We do not fit the photometry in this analysis. We follow an iterative approach where we add a Keplerian signal to the model at a time using wide, uninformative priors on all model parameters \citep[see][for a review of statistical methods for analyzing RV datasets]{HaraFord2023AnRSA..10..623H}. Results are reported in Table~\ref{tab:model_comparison}. 

First, we find that, as in \citet{Martioli2023A&A...680A..84M}, models with or without fitting for the eccentricity are statistically indistinguishable ($\Delta\ln Z < 1$). Thus, we assume circular orbits for all the signals in this analysis from here onward. Then, we find that adding a second signal to the model with an uninformative wide prior between 1 and 1000\,days, results in a much better fit ($\Delta\ln Z \sim 4$). Different instances of that model result in a posterior distribution on the period centered at either 5.46 or 60.45\,days, but there is no preference for including one over the other. But, the best-fit to the data is with a model that includes all three signals ($\Delta\ln Z = 15.7$ over a 1p model, but $>11$ over 2p models). 

This 3p model is preferred even in the presence of correlated noise, which we model using a Gaussian Process (GP) exponential sine-squared kernel from \texttt{george} of the form
\begin{equation}
    k_{i,j} = \sigma^2 \exp\left(- \alpha (t_i - t_j)^2 - \Gamma \sin^2 \left[\frac{\pi |t_i - t_j|}{P_{\rm rot}}\right]\right) \qquad .
\end{equation}
Testing with other GP kernels (exponential decay or quasi-periodic) give similar results ($\Delta\ln Z \sim 0.2$). We set log-uniform priors for $\sigma$ between 0.1 and 300\,$\mathrm{m\,s^{-1}}$, $\alpha$ between $10^{-8}$ and $10^{-3}$, $\Gamma$ between 0.01 and 10, and a uniform prior for $P_{\rm rot}$ between 10 and 70\,days. The choice of the prior on $P_{\rm rot}$ is motivated to try to distinguish if the 60-day signal is of stellar origin, although the typical rotation period of Sun-like stars is between 10 and 30 days \citep[e.g.,][]{McQuillan2014ApJS..211...24M} and would be unexpected. Furthermore, none of the activity indicators from the HARPS-N spectra show hints of stellar variability at that or any other period (Fig.~\ref{fig:gls_activity}). The hyperparameters from these GP models show that $P_{\rm rot} = 60.5\pm0.6\,\mathrm{d}$ and $\alpha \sim 10^{-7}$, which indicates that the GP is modeling a signal close to be perfectly periodic with a very well-defined frequency unlikely to be of stellar origin \citep{Stock2023A&A...674A.108S}. 

We find that although a 2p model (with the transiting planet and the 5.46-day signal) and a GP explain the data well, assuming a circular orbit for the 60-day signal is not only a simpler model but statistically preferred ($\Delta\ln Z = 4.6$). A stacked Bayesian GLS periodogram analysis following \citet{SBGLS} reveals that all three significant RV signals (at 5.46, 18.26, and 60 days) are coherent and stable throughout the observing period, reinforcing the notion that they are of planetary nature (Fig.~\ref{fig:sbgls}). Besides, it is important to highlight that the derived semi-amplitude for the transiting planet ($K_{\rm b}$) remains within 1$\sigma$ for every model tested, ensuring a robust mass determination of the planet regardless of the chosen model. 

To conclude, we find that the model with the highest statistical evidence is a three-planet model (assuming circular orbits) with no need to account for correlated noise. There is enough statistical evidence to claim the 5.46- and 60-day signals as present (and significant) in the RV data and of likely planetary origin, in line with other works that follow-up transiting planets with RV instruments and find internal and/or external companions \citep[e.g.,][]{Luque2019,Kemmer2020}.

\subsection{Joint fit} \label{subsec:joint_fit}

\begin{figure*}[ht!]
    \centering
    \includegraphics[width=\hsize]{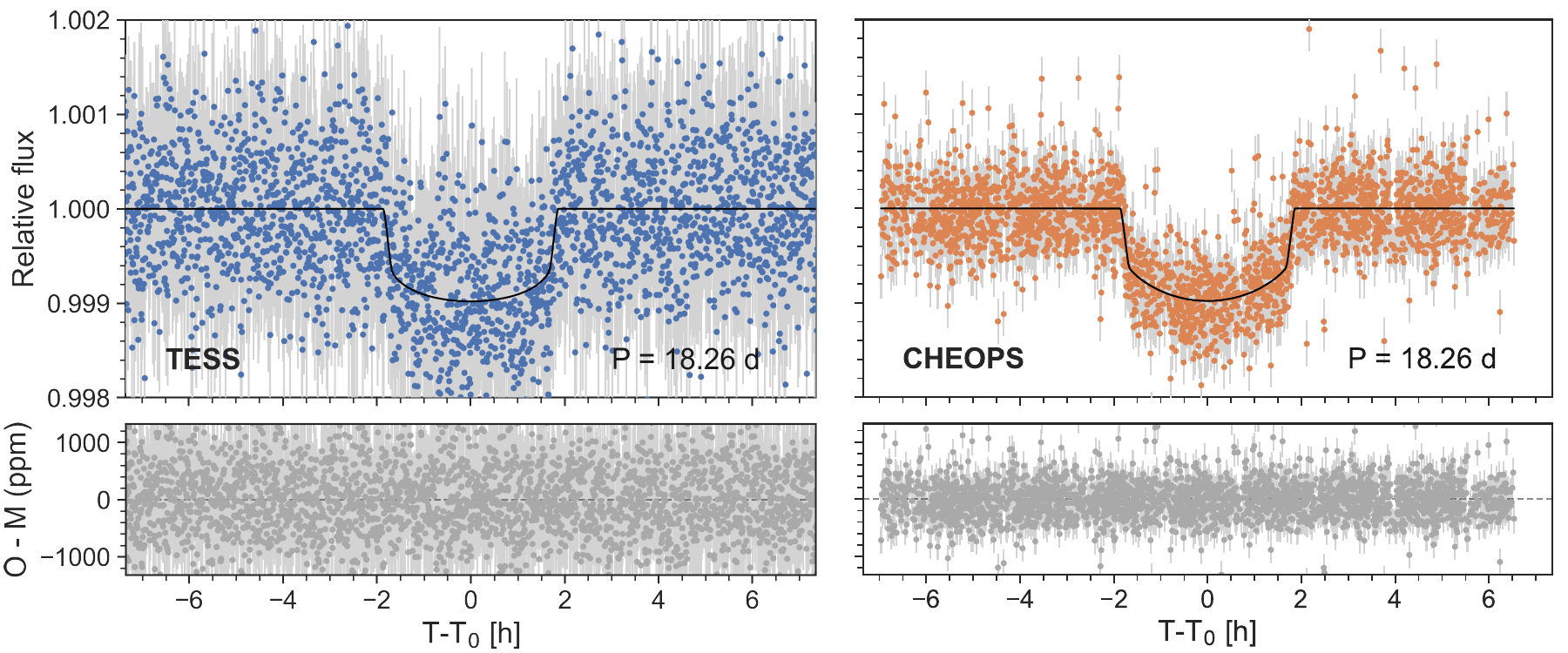}\\
    \includegraphics[width=\hsize]{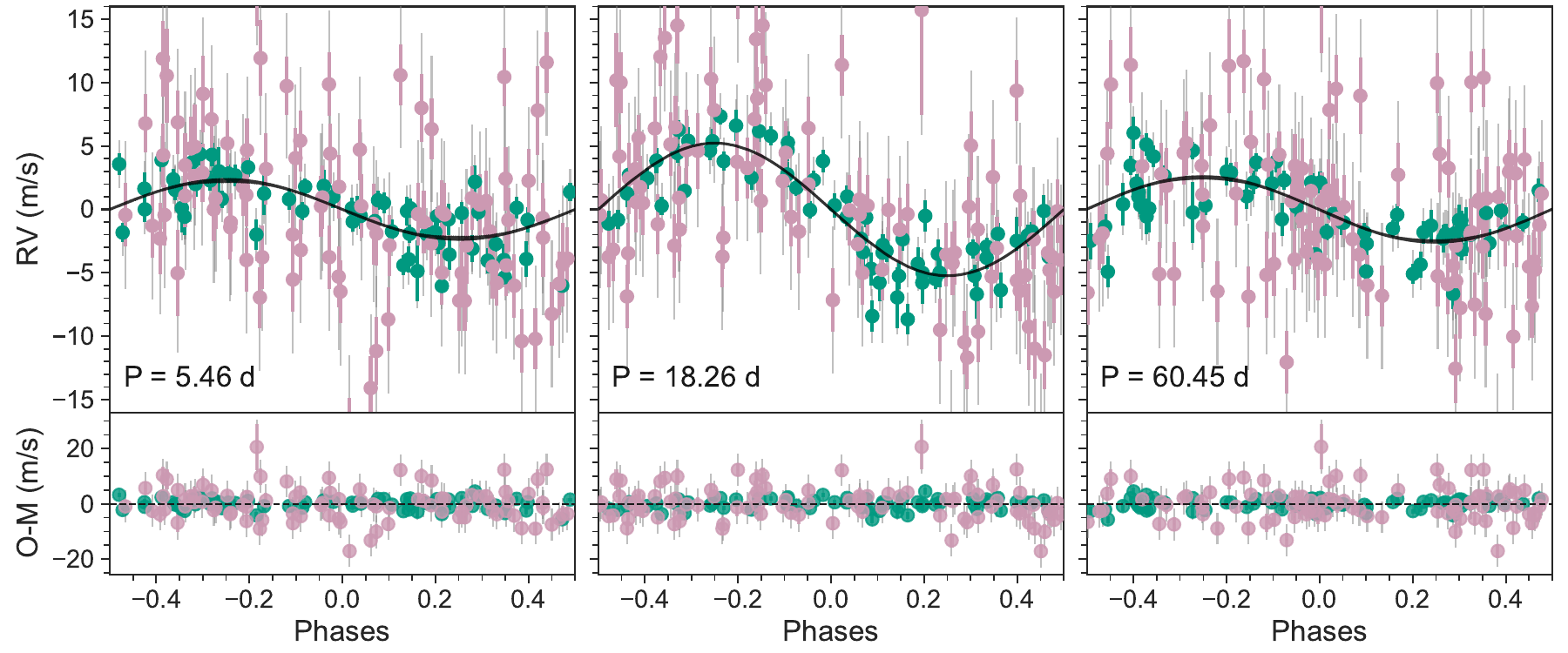}
    \caption{\textit{Top}: TESS (left) and CHEOPS (right) photometry phase-folded to the 18.26-day period of the transiting planet TOI-2141~b. \textit{Bottom}: RVs phase-folded to the period of all planets (TOI-2141~c, left; TOI-2141~b, center; TOI-2141~d, right). Residuals are shown in the bottom of each panel. In all panels, the black line is the best-fit model from the joint fit described in Sect.~\ref{subsec:joint_fit}.}
    \label{fig:joint_planets}
\end{figure*}

\renewcommand{\arraystretch}{1.4}
\begin{table*}
    \centering
    \caption{Priors and posterior distributions (median and 68\% credibility intervals) for each fit parameter of the final joint model obtained for the TOI-2141 system using \texttt{juliet}. }
    \label{tab:posteriors}
\begin{tabular}{llcr}
\hline \hline
Parameter & Prior & Posterior & Units \\
\hline
$\rho_\star$        & $\mathcal{N}(1473, 100)$ & ${1521}^{+{96}}_{-{95}}$ & \si{\kilo\gram\per\meter\cubed} \\
\hline
$P_\mathrm{c}$      & $\mathcal{U}(4.8, 6.0)$ & ${5.4624}^{+{0.0026}}_{-{0.0027}}$ & d \\
$t_{0,\mathrm{c}}$  & $\mathcal{U}(1976.0, 1982.0)$ & ${1979.19}^{+{0.58}}_{-{0.56}}$ & BJD - 2457000 \\
$K_\mathrm{c}$      & $\mathcal{U}(0, 20)$ & ${2.40}^{+{0.33}}_{-{0.33}}$ & $\mathrm{m\,s^{-1}}$ \\
\hline
$P_\mathrm{b}$      & $\mathcal{N}(18.26,0.1)$ & ${18.261608}^{+{1.9e-05}}_{-{2e-05}}$ & d \\
$t_{0,\mathrm{b}}$  & $\mathcal{N}(1992.5, 1.0)$ & ${1992.502}^{+{0.0014}}_{-{0.0013}}$ & d \\
$r_{1,\mathrm{b}}$  & $\mathcal{U}(0, 1)$ & ${0.767}^{+{0.013}}_{-{0.014}}$ & \dots \\
$r_{2,\mathrm{b}}$  & $\mathcal{U}(0, 1)$ & ${0.03037}^{+{0.00035}}_{-{0.00034}}$ & \dots \\
$K_\mathrm{b}$      & $\mathcal{U}(0, 20)$ & ${5.25}^{+{0.35}}_{-{0.35}}$ & $\mathrm{m\,s^{-1}}$ \\
\hline
$P_\mathrm{d}$      & $\mathcal{U}(50, 70)$ & ${60.45}^{+{0.24}}_{-{0.25}}$ & d \\
$t_{0,\mathrm{d}}$  & $\mathcal{U}(1900.0, 1960.0)$ & ${1940.5}^{+{5.2}}_{-{5.2}}$ & d \\
$K_\mathrm{d}$      & $\mathcal{U}(0, 20)$ & ${2.61}^{+{0.34}}_{-{0.34}}$ & $\mathrm{m\,s^{-1}}$ \\
\hline
$\gamma_{\mathrm{HARPS-N}}$ & $\mathcal{U}(-19900.0, -19800.0)$ & ${-19829.97}^{+{0.26}}_{-{0.25}}$ & $\mathrm{m\,s^{-1}}$ \\
$\sigma_{\mathrm{HARPS-N}}$ & $\mathcal{J}(0.1, 20)$ & ${1.70}^{+{0.23}}_{-{0.30}}$ & $\mathrm{m\,s^{-1}}$ \\
$\gamma_{\mathrm{SOPHIE}}$  & $\mathcal{U}(-19900.0, -19800.0)$ & ${-19860.58}^{+{0.62}}_{-{0.60}}$ & $\mathrm{m\,s^{-1}}$ \\
$\sigma_{\mathrm{SOPHIE}}$  & $\mathcal{J}(0.1, 20)$ & ${5.25}^{+{0.62}}_{-{0.60}}$ & $\mathrm{m\,s^{-1}}$ \\
$q_{1,\mathrm{TESS}}$       & $\mathcal{U}(0, 1)$ & ${0.30}^{+{0.18}}_{-{0.13}}$ & \dots \\
$q_{2,\mathrm{TESS}}$       & $\mathcal{U}(0, 1)$ & ${0.37}^{+{0.34}}_{-{0.25}}$ & \dots \\
$q_{1,\mathrm{CHEOPS}}$     & $\mathcal{U}(0, 1)$ & ${0.32}^{+{0.08}}_{-{0.06}}$ & \dots \\
$q_{2,\mathrm{CHEOPS}}$     & $\mathcal{U}(0, 1)$ & ${0.76}^{+{0.17}}_{-{0.24}}$ & \dots \\
$\sigma_{\mathrm{TESS}}$    & $\mathcal{J}(0.1, 1000)$ & ${1.34}^{+{6.52}}_{-{1.11}}$ & ppm \\
$\sigma_{\mathrm{CHEOPS}}$  & $\mathcal{J}(0.1, 1000)$ & ${230.85}^{+{6.65}}_{-{6.40}}$ & ppm \\
\hline
\end{tabular}
\tablefoot{
    The prior labels of $\mathcal{N}$, $\mathcal{U}$, and $\mathcal{J}$ represent normal, uniform, and Jeffrey's distributions, respectively.
}
\end{table*}

\renewcommand{\arraystretch}{1.4}
\begin{table*}
    \centering
    \caption{Derived planetary parameters obtained for the TOI~2141 system using the posterior values from the joint fit in Table~\ref{tab:posteriors} and stellar parameters from Table~\ref{tab:star}. }
    \label{table:params_derived}
\begin{tabular}{ccccc}
\hline \hline
Parameter & Planet c & Planet b & Planet d & Units \\
\hline
$p = R_\text{p}/R_\star$ & \dots & ${0.03037}^{+{0.00035}}_{-{0.00034}}$ & \dots & \dots \\
$b = (a_\text{p}/R_\star)\cos i_\text{p}$ & \dots & ${0.65}^{+{0.019}}_{-{0.02}}$ & \dots & \dots \\ 
$a_\text{p}/R_\star$ & ${13.39}^{+{0.27}}_{-{0.28}}$ & ${29.93}^{+{0.62}}_{-{0.64}}$ & ${66.47}^{+{1.37}}_{-{1.43}}$ & \dots \\
$i_\text{p}$ & \dots & ${88.755}^{+{0.063}}_{-{0.063}}$ & \dots & deg \\
$M_\text{p}$ & \dots & ${20.1}^{+{1.6}}_{-{1.5}}$ & \dots & $M_\oplus$ \\
$M_\text{p}\sin i$ & ${6.14}^{+{0.86}}_{-{0.86}}$ & ${20.1}^{+{1.6}}_{-{1.5}}$ & ${14.9}^{+{2.1}}_{-{2.1}}$ & $M_\oplus$ \\
$R_\text{p}$ & \dots & ${3.147}^{+{0.043}}_{-{0.042}}$ & \dots & $R_\oplus$ \\
$\rho_\text{p}$ & \dots & ${3.54}^{+{0.31}}_{-{0.3}}$ & \dots & \si{\gram\per\centi\meter\cubed} \\
$g_\text{p}$ & \dots & ${19.9}^{+{1.7}}_{-{1.6}}$ & \dots & \si{\meter\per\second\squared} \\
$a_\text{p}$ & ${0.0585}^{+{0.0011}}_{-{0.0012}}$ & ${0.1322}^{+{0.003}}_{-{0.003}}$ & ${0.2906}^{+{0.0055}}_{-{0.0057}}$ & \si{\astronomicalunit} \\
$T_\textnormal{eq, p}$\tablefootmark{(a)} & ${1095.0}^{+{17.0}}_{-{17.0}}$ & ${728.0}^{+{11.0}}_{-{11.0}}$ & ${491.4}^{+{7.4}}_{-{7.2}}$ & \si{\kelvin} \\
$S$ & ${241.0}^{+{12.0}}_{-{11.0}}$ & ${47.3}^{+{2.5}}_{-{2.4}}$ & ${9.8}^{+{0.47}}_{-{0.43}}$ & $S_\oplus$ \\
TSM & \dots & ${35.6}^{+{3.3}}_{-{2.9}}$ & \dots & \dots \\
\hline
\end{tabular}
\tablefoot{\tablefoottext{a}{Assuming zero Bond albedo and perfect energy redistribution.}}
\end{table*}

Using \texttt{juliet}, we jointly fit the TESS, CHEOPS, SOPHIE, and HARPS-N data. To speed up computational times, we use the TESS PDC photometry without additional detrending and the detrended CHEOPS data analyzed in Sect.~\ref{subsec:CHEOPS_fit}. For the model, we assume the configuration inferred from our Bayesian model comparison RV-only analysis presented in Sect.~\ref{subsec:rv_fit} and shown in Table~\ref{tab:model_comparison}. For the photometry part of our fit, we adopted a quadratic limb darkening law for TESS and CHEOPS data parameterized by the coefficients $q_1, q_2$ introduced by \citet{Kipping13} and fit them as free parameters with unconstrained priors. The transiting planet is modeled using the $r_1$, $r_2$ parameterization of the planet-to-star radius ratio $p=R_p/R_\star$ and the impact parameter of the orbit $b$ introduced by \citet{Espinoza18}. Additionally, rather than fitting the scaled planetary radius ($a/R_\star$) for each planet, we use the stellar density ($\rho_\star$) as a free parameter and set a normal prior based on the stellar properties described in Table~\ref{tab:star}. We added a jitter term $\sigma$ to the photometric uncertainties of TESS and CHEOPS data. For the radial velocity part of our fit, we assumed circular orbits for all planets. The priors on the period and phase of each planet were informed by our previous transit- and RV-only analyses, but wide enough to derive reliable uncertainties. We add an instrumental offset and a jitter term for each RV instrument and no GP. The model parameters, their priors, and posterior distributions are reported in Table~\ref{tab:posteriors}. The best-fit models are shown in Figs.~\ref{fig:tess_lc}~and~\ref{fig:rvs} for the photometry and RV timeseries, and phase-folded to the periods of interest in Fig.~\ref{fig:joint_planets}.

\section{Discussion} \label{sec:discussion}

\begin{figure}[ht!]
    \centering
    \includegraphics[width=\hsize]{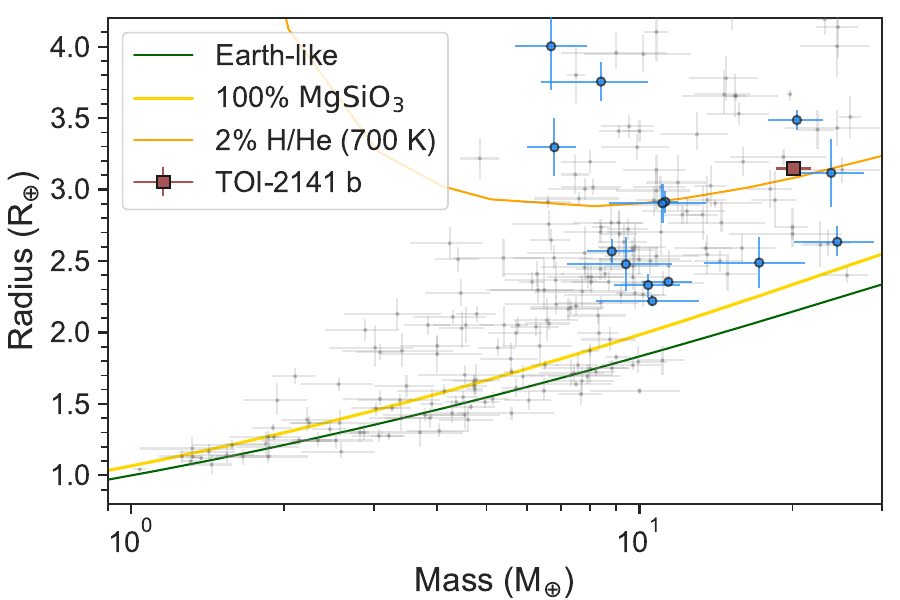}
    \caption{Mass-radius diagram of transiting planets in the PlanetS catalog \citep{Otegi2020,Parc2024} as of May 2025. Internal structure models from \citet{Zeng2019}. The location of TOI-2141~b is marked in red. For this planet, the marker size is larger than the uncertainty in the y-axis. Transiting planets with orbital periods longer than 15\,days orbiting Sun-like stars ($5500\,\mathrm{K} < T_{\rm eff} < 5900\,\mathrm{K}$) are shown in blue. Compared to similar planets, the precision in the bulk properties of TOI-2141~b is remarkable.}
    \label{fig:mr_diagram}
\end{figure}

Table~\ref{table:params_derived} shows the fundamental physical parameters derived for the three planets based on the model posteriors from Table~\ref{tab:posteriors} and the stellar properties from Table~\ref{tab:star}. In summary, we refine the radius and mass determination of the known transiting planet TOI-2141~b (with 1.3\% and 7.9\% radius and mass uncertainties, respectively\footnote{For comparison, the precision in radius and mass reported in \citet{Martioli2023A&A...680A..84M} is of 7.5\% and 16.7\%, respectively.}) and report the discovery of an inner and an outer non-transiting companion to the known planet. 

Compared to \citet{Martioli2023A&A...680A..84M}, we find that TOI-2141~b is slightly larger (3.147 vs. $3.05\,R_\oplus$) and less massive (20.1 vs. $24\,M_\oplus$), making the planet puffier (3.5 vs. $4.6\,\mathrm{g\,cm^{-3}}$) although all differences are within the $1\sigma$ uncertainties reported in that work. Figure~\ref{fig:mr_diagram} shows the position of the planet in a mass-radius diagram with respect to other transiting planets orbiting Sun-like stars at relatively long-period orbits. We note that the precision on the bulk properties of TOI-2141~b is significantly higher than that of the comparison population and that the planet is still relatively dense compared to similar sub-Neptunes.

\subsection{Interior structure and atmospheric evolution of TOI-2141~b} \label{subsec:interior_atmevol}

\begin{figure*}[t]
    \centering
    \includegraphics[width=\linewidth]{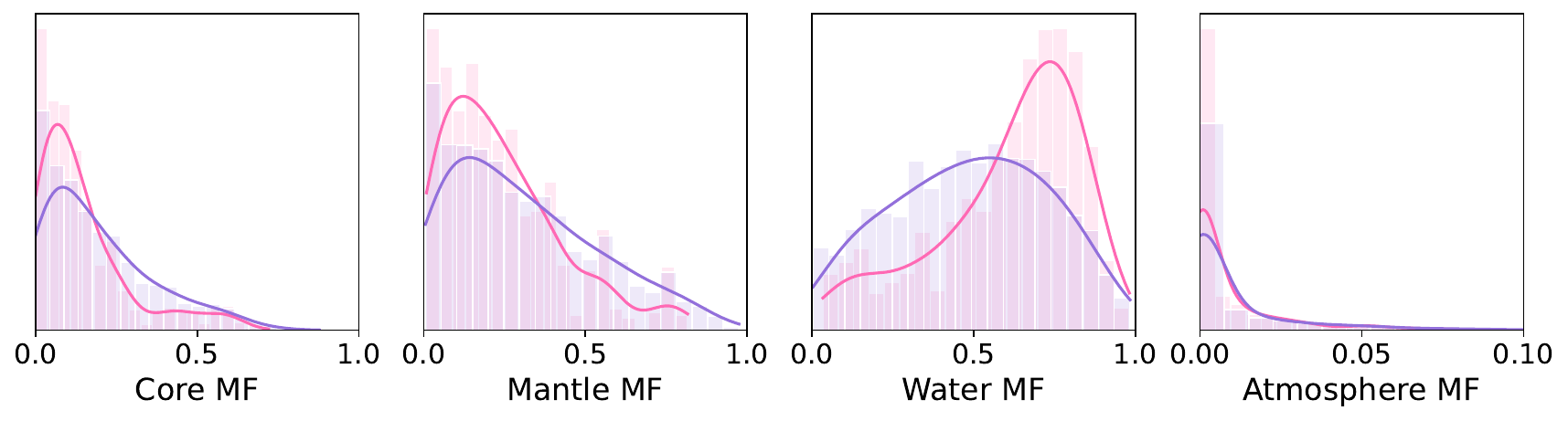}
    \caption{ExoMDN computed mass fractions of the core, mantle, water and atmosphere layers for the our derived planetary parameters in pink and the planetary parameters derived by \citet{Martioli2023A&A...680A..84M} in purple.}
    \label{fig:exomdn}
\end{figure*}

\begin{figure}[t]
    \centering
    \includegraphics[width=\hsize]{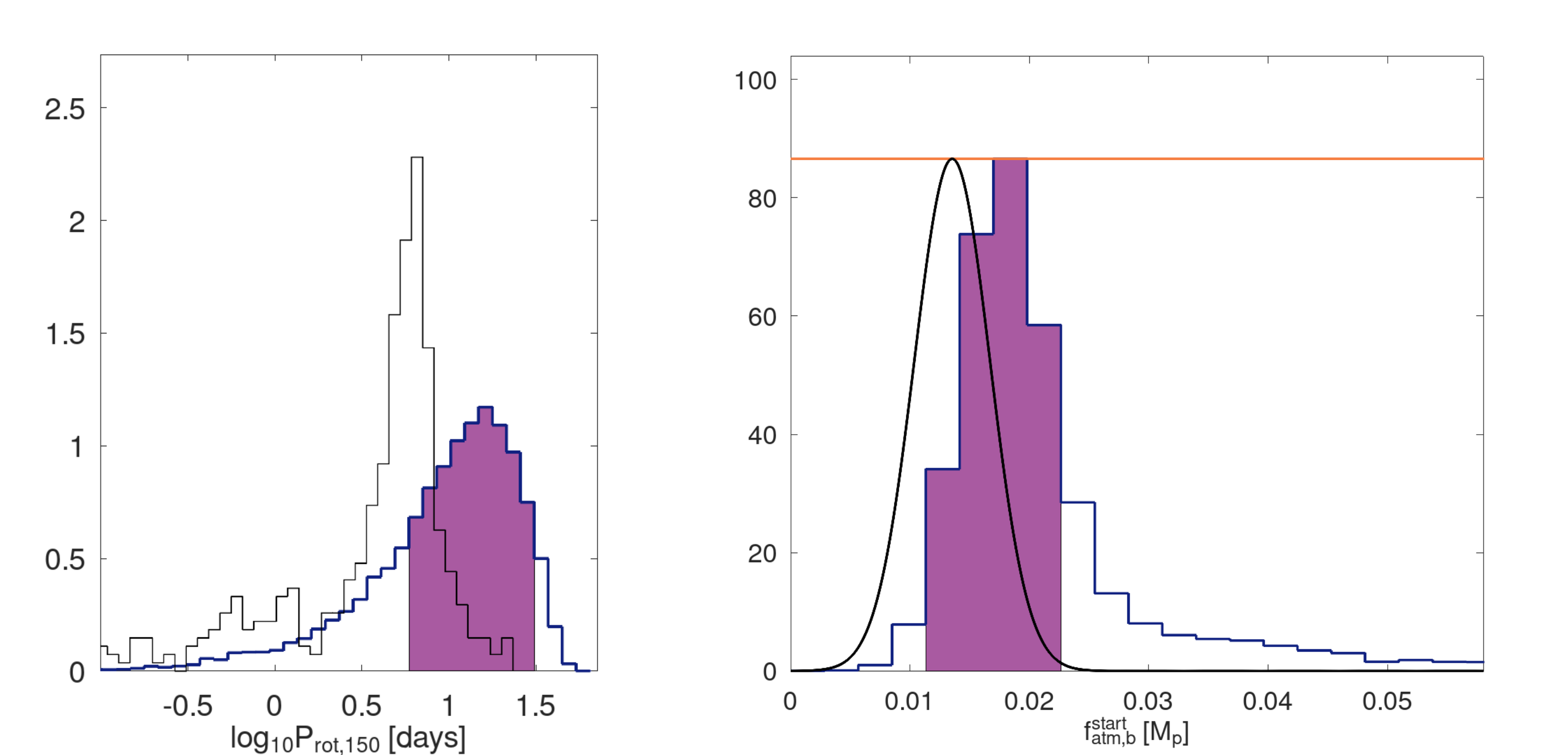} \\
    \caption{Posterior distributions (pdf) as derived from \texttt{PASTA} (blue histograms), with the violet area that highlights the 68.3\% highest probability density interval. \textit{Left panel:} Rotation period pdf of TOI-2141 when the star was 150 Myr old ($P_{\mathrm{rot,150}}$) compared with the $P_{\mathrm{rot,150}}$ distribution of stars of comparable masses (black histogram) as extracted from the sample by \citet{johnstone2015Prot150}. \textit{Right panel:} Initial atmospheric mass fraction ($f_{\mathrm{atm}}^{\mathrm{start}}$) pdf of TOI-2141\,b compared with the model-inferred present-day atmospheric content (black line). The orange horizontal line marks the uniform prior that was imposed on $f_{\mathrm{atm}}^{\mathrm{start}}$.}
    \label{fig:fatm_Prot}
\end{figure}

We use the publicly available tool \texttt{ExoMDN} \citep{Baumeister_ExoMDN_2023}, which uses machine learning techniques, particularly mixture density networks, to characterize planets quickly, to model the interior structure of TOI-2141~b in more detail. The code is trained on $\sim$5.6 million synthetic planets within mass ranges from 0.1--25\,M$_\oplus$ and equilibrium temperatures from 100--1000\,K. \texttt{ExoMDN} computes the interior structure of a planet using \texttt{TATOOINE} \citep{Baumeister_TATOOINE_2020} and assumes that the planet consists of four distinct layers, which is a strong simplification for sub-Neptune planets \citep[see e.g.,][]{Luo2024}. The centre of the planet is made up of a core consisting of solid iron. This core is surrounded by the second layer, a silicate mantle divided into an upper and lower mantle, assumed to have an Earth-like composition. On top of this is a water layer, which follows the equations of state from \texttt{AQUA} \citep{Haldemann_AQUA_2020}. Finally, the outer layer is an atmosphere assumed to be made up of hydrogen and helium with solar ratios. The atmosphere is treated as isothermal in the model at the equilibrium temperature of the planet.

We use the derived radius, mass, and equilibrium temperature from Table~\ref{table:params_derived} as our input and use the default sample of 1000 uncertainty samples to predict the interior structure distribution for inputs within the error bars. We compute 10 samples of each distribution to reach a total sample of 10,000. The resulting mass fractions are shown in pink in Fig.~\ref{fig:exomdn}. \texttt{ExoMDN} predicts a core mass fraction of $0.103_{-0.077}^{+0.144}$, and a mantle mass fraction of $0.20_{-0.14}^{+0.21}$ while the water mass fraction lies at $0.67_{-0.32}^{+0.14}$ and the atmospheric mass fraction of $0.00039_{-0.00039}^{+0.01636}$. These errors may be underestimated given that the errors of an individual sample are from the search algorithm, not accounting for physical uncertainties.

We apply \texttt{ExoMDN} in the same way to model the interior structure with the planetary parameters derived by \citet{Martioli2023A&A...680A..84M} as input. This results in a core mass fraction of $0.15_{-0.11}^{+0.22}$, a mantle mass fraction of $0.26_{-0.19}^{+0.30}$, water mass fraction of $0.50_{-0.29}^{+0.25}$ and an atmospheric mass fraction of $0.00024_{-0.00024}^{+0.01743}$. These values are in agreement with the interior structure derived from the planetary parameters derived in this work. A comparison between the two is shown in Fig.~\ref{fig:exomdn}, where the pink line represents the results from our work, while the purple line is the \texttt{ExoMDN} results of the planetary parameters from \citet{Martioli2023A&A...680A..84M}. This shows that our newly derived radius and mass allow us to better constrain the layers. 

In addition, we modeled the atmospheric evolution of TOI-2141~b using the \textsl{P}lanetary \textsl{A}tmosphere and \textsl{S}tellar Ro\textsl{T}ation R\textsl{A}te \citep[\texttt{PASTA};][]{bonfanti2021b} code, which works in an MCMC fashion within the \texttt{MC3} Bayesian framework developed by \citet{cubillos2017}. The algorithm employs MESA evolutionary tracks \citep{choi2016}, a gyrochronological relation in the form of a broken power-law \citep{bonfanti2021b}, an empirical model to relate stellar rotation with the X-ray and extreme ultraviolet (XUV) flux \citep[][and references therein]{bonfanti2021b}, a planetary structure model that links planetary observables to the atmospheric mass \citep{johnstone2015planetModels}, and a model for computing the atmospheric escape rate, which results from internal heating and XUV stellar irradiation \citep{kubyshkina2018aMdot,kubyshkina2018bMdot}. \texttt{PASTA} neglects any migration that might have occurred after the proto-planetary disk dispersal and assumes that the atmospheric content of the planets, if any, is given by H-dominated atmospheres.

At each MCMC step, \texttt{PASTA} computes the evolutionary track of the exoplanet atmospheric mass fraction $f_{\mathrm{atm}}(t)$ and the step is accepted if the present-day planet radius inferred from $f_{\mathrm{atm}}(t_{\star})$ is consistent with its observational counterpart. As a result, we obtained the posterior distribution for the initial atmospheric mass fraction $f_{\mathrm{atm}}^{\mathrm{start}}\equiv f_{\mathrm{atm}}(t_{\mathrm{disk}})$ of TOI-2141 b (blue histogram in the second panel of Fig.~\ref{fig:fatm_Prot}), where $t_{\mathrm{disk}}$ is the time of dispersal of protoplanetary disk. According to our framework, TOI-2141\,b has lost an atmospheric content roughly equal to 1\% of its total mass during its lifetime. By modelling the stellar activiy, \texttt{PASTA} also outputs the expected stellar rotation period at $t=150$\,Myr and compare it with the period distribution of coeval stars having masses comparable to TOI-2141 as extracted from \citet{johnstone2015Prot150} (thick blue and thin black histograms in the first panel of Fig.~\ref{fig:fatm_Prot}, respectively), from which we infer that the star was likely born as a slow rotator.

\subsection{Do the RV-only planets transit?} \label{subsec:transit_c}

This is a sensible question given that the innermost planet in the system, TOI-2141~c, has a transit probability of $7.5\pm0.1$\% using the values in Table~\ref{table:params_derived} (and 100\% if the orbits would be perfectly coplanar). For comparison, planet b has a 3.3\% transit probability and d just 1.5\%. To answer it, we carried out a joint fit as in Sect.~\ref{subsec:joint_fit}, but assuming that all three planets are transiting. Our priors were thus identical as in Table~\ref{tab:posteriors}, but adding the parameters $r_1$ and $r_2$ for both planets c and d. The posteriors from this analysis are virtually identical to those shown in Table~\ref{tab:posteriors} except for those parameters and are thus not reported in this manuscript. 

Figure~\ref{fig:nontransiting} shows the TESS photometry phase-folded to the periods of the c and d planets from that joint fit. There is no evidence for either of the planets to be transiting during the time of TESS observations. The CHEOPS data, scheduled to coincide with transits of planet b, do not cover either of the two planets' predicted transit windows. For the innermost planet, this is surprising given the closeness to the star, but a model that assumes the planet to be transiting is significantly disfavored ($\Delta\ln Z = 351679.2 - 351693.6 = -14.4$). If planet c would transit, our joint fit constrain its size to be smaller than ${0.34}^{+{0.32}}_{-{0.23}}\,R_\oplus$, which combined with its inferred mass of $\sim 6\,M_\oplus$ would imply a density of $850\,\si{\gram\per\centi\meter\cubed}$, more than a hundred times denser than iron. For planet d, the inferred density would be even larger as the planet is more massive than c although its long period makes it more likely to be non-transiting from our line of sight. The minimum mutual inclination between planets b and c given these constrains $\Delta i_{\text{min}} \approx \cos^{-1}\left( \frac{R_\star}{a_b} \right) - \cos^{-1}\left( \frac{R_\star}{a_c} \right) \geq 2.4\,\deg$.

\subsection{System stability and dynamical constrains} \label{subsec:dynamics}

\begin{figure}
    \centering
	\includegraphics[width=0.5\textwidth]{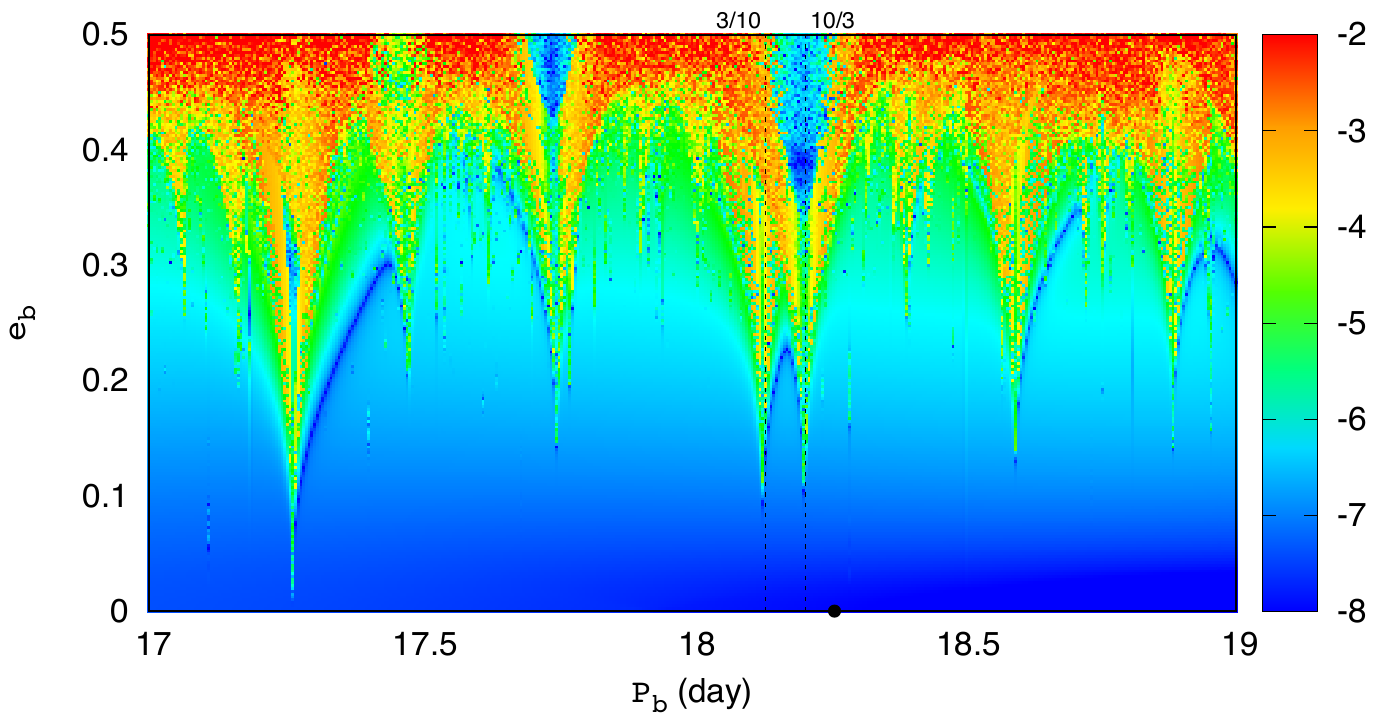}
    \caption{Stability analysis of the TOI-2141 planetary system in the plane $(P_b, e_b)$, assuming coplanar orbits. For fixed initial conditions, the parameter space of the system is explored by varying the orbital period and the eccentricity of planet-$b$. The step size is $0.005$~day in the orbital period and $0.0025$ in the eccentricity. For each initial condition, the system is integrated over $10^4$~yr, and a stability indicator is calculated, which involves a frequency analysis of the mean longitude of planet b. The chaotic diffusion is measured by the variation in the frequency (see text). Red points correspond to highly unstable orbits, while blue points correspond to orbits that are likely to be stable on Gyr timescales. The vertical dashed lines correspond to both 10/3 period ratios with planets c and d. The black dot shows the best fit solution from Sect.~\ref{subsec:joint_fit}. }
    \label{figstab1}
\end{figure}

\begin{figure*}
    \centering
	\includegraphics[width=0.49\textwidth]{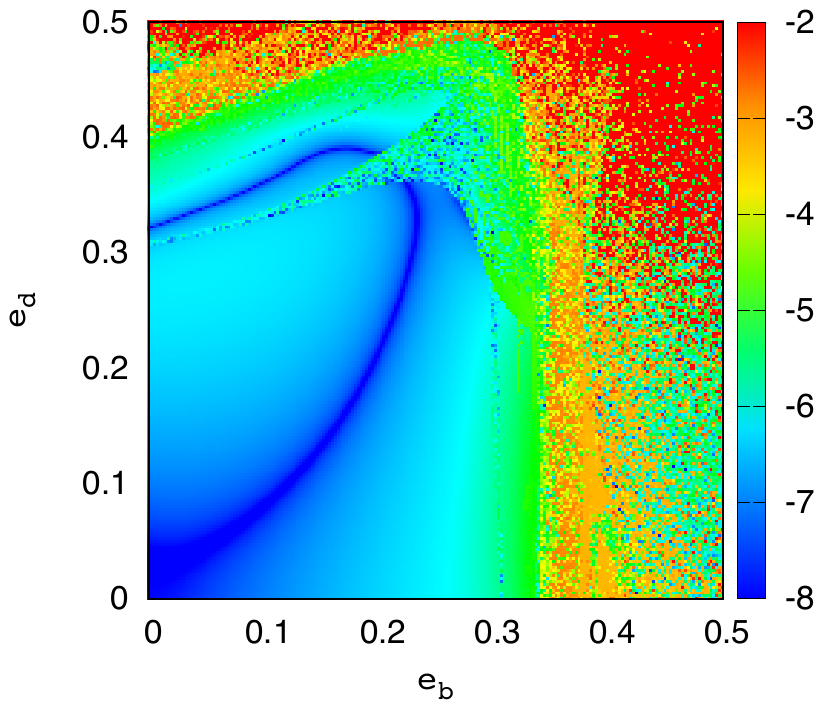}
    \includegraphics[width=0.49\textwidth]{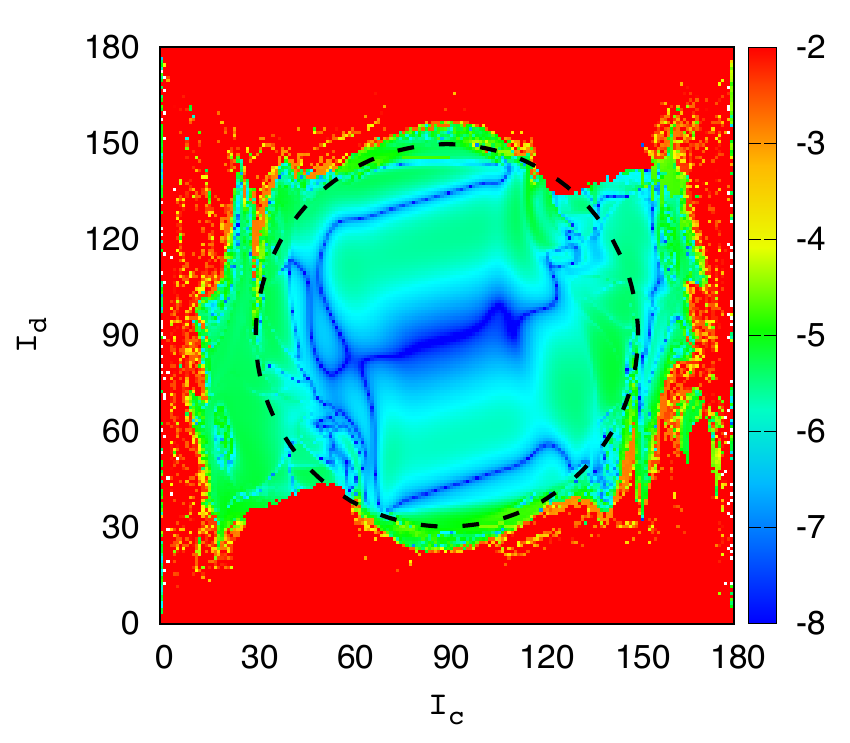}
    \caption{\textit{Left:} Stability analysis in the plane $(e_b, e_d)$, assuming coplanar orbits. For fixed initial conditions, the parameter space of the system is explored by varying the eccentricities of planets b and d with a step size of $0.0025$. \textit{Right:} Stability analysis in the plane $(i_c, i_d)$, assuming $i_b = 88.76$\,deg (Table~\ref{table:params_derived}) and the longitude of the node of all planets equal to zero. For fixed initial conditions, the parameter space of the system is explored by varying the inclinations of planets c and d with a step size of 1\,deg. The dashed circle corresponds to mutual inclinations with planet b of 60\,deg. The color codes are the same as in Fig.~\ref{figstab1}.}
    \label{figstab2}
\end{figure*}

The TOI-2141 system hosts three sub-Neptune-mass planets in relatively well-spaced orbits. Interestingly, the system seems to be close to an unusual $10/3$ mean motion resonance chain, so the stability is not straightforward or warranted. To get a clear view on the stability of the system, we performed a dynamical analysis as in \citet{Correia_etal_2005, Correia_etal_2010}. The system is integrated on a regular 2D mesh of initial conditions around the best fit (Table~\ref{tab:posteriors}). Each initial condition is integrated for $10^4$~yr, using the symplectic integrator \texttt{SABAC4} \citep{Laskar_Robutel_2001}, with a step size of $2.5 \times 10^{-4} $~yr and general relativity corrections. Then, we perform a frequency analysis \citep{Laskar_1990, Laskar_1993PD} of the mean longitude of planet b over two consecutive time intervals of 5000~yr, and determine the main frequency, $n$ and $n'$, respectively. The stability is measured by $\Delta = |1-n'/n|$, which estimates the chaotic diffusion of the orbits. The stability results are visually reported as color in Figs.~\ref{figstab1} and \ref{figstab2}. Orange and red represent chaotic, strongly unstable trajectories; yellow indicates the transition between stable and unstable regimes; green corresponds to moderately chaotic trajectories, but stable on Gyr timescales; cyan and blue give extremely stable quasi-periodic orbits.

In Fig.~\ref{figstab1}, we explore the stability of planet b by varying its orbital period and eccentricity. We note that although the orbital period of planet b is very well constrained, we opted to vary this parameter such that we can simultaneously view both 10/3 period ratios with planets c and d (marked with dashed lines). We observe that the best-fit solution (black dot) is close, but outside both resonances, even if we include the observational errors. Thus, the resonant motion can be discarded. Besides, the best fit is embedded in a large stability region (blue), so we can conclude that the orbits of all planets are trustworthy from a stability point of view. 

In Fig.~\ref{figstab2}, we explore the stability of the system by varying the eccentricities of planets b and d --- the two most massive in the system. Although in the best-fit solution their eccentricities were fixed at zero, we observe that the system is resilient even for high values, up to about 0.3. We vary only $e_{\rm b}$ and $e_{\rm d}$, because planet c is the closest to the star, and so it is more likely that $e_{\rm c} \sim 0$ owing to tidal damping.

Figure~\ref{figstab2} shows the inclinations to the line-of-sight of planets c and d. We vary only $i_{\rm c}$ and $i_{\rm d}$, because planet b is transiting, and thus its inclination is well determined, $i_b = 88.76\pm0.06\,\mathrm{deg}$ (Table~\ref{table:params_derived}). For guidance, we also show a dashed circle that corresponds to mutual inclinations with planet b of 60\,deg. We observe that planet c can be stable for $i_c > 15$\,deg, so its maximal mass is about $24\,M_\oplus$, while planet d can be stable for $i_d > 30$\,deg, so its maximal mass is about $30\,M_\oplus$. This analysis shows all planets in the system, despite two being non-transiting, have masses below $30\,M_\oplus$ from stability arguments.

\section{Conclusions} \label{sec:conclusions}

We refined the characterization of the TOI-2141 system using new TESS and CHEOPS photometry, along with 61 HARPS-N RVs. For the known transiting planet TOI-2141~b, we determine a radius of $3.15\pm0.04\,R_\oplus$ and mass of $20.1\pm1.6\,M_\oplus$, corresponding to a bulk density of $3.54\pm0.31\,\mathrm{g\,cm^{-3}}$. Interior structure models suggest a substantial water-rich layer and a thin hydrogen-dominated atmosphere, but stronger conclusions cannot be made without future observations focusing on the atmospheric composition. The high precision achieved in the planetary mass and radius of TOI-2141~b (7.9\% and 1.3\%, respectively) makes it a valuable benchmark for studies of sub-Neptune composition and evolution.

Bayesian model comparison analyses on the RV data strongly support a three-planet model, with additional Keplerian signals at 5.46 and 60.45\,days. Despite the high transit probability of the inner companion, a joint analysis rules out transits of planets c and d with high confidence. This implies a minimum mutual inclination of $\geq2.4^\circ$ between planets b and c, suggesting a slightly misaligned but dynamically quiet system. This system adds to the growing evidence that many single-transiting planets are part of multi-planet systems with small but significant inclinations.

\begin{acknowledgements}
R.L. is funded by the European Union (ERC, THIRSTEE, 101164189). Views and opinions expressed are however those of the author(s) only and do not necessarily reflect those of the European Union or the European Research Council. Neither the European Union nor the granting authority can be held responsible for them. R.L. is supported by NASA through the NASA Hubble Fellowship grant HST-HF2-51559.001-A awarded by the Space Telescope Science Institute, which is operated by the Association of Universities for Research in Astronomy, Inc., for NASA, under contract NAS5-26555. R.L. acknowledges financial support from the Severo Ochoa grant CEX2021-001131-S funded by MCIN/AEI/10.13039/501100011033.

Data for this paper has been obtained under the International Time Programme of the CCI (International Scientific Committee of the Observatorios de Canarias of the IAC) with the Telescopio Nazionale Galileo operated on the island of La Palma by the Fundacion Galileo Galilei - INAF, Fundacion Canaria in the Observatorio Roque de los Muchachos.

CHEOPS is an ESA mission in partnership with Switzerland with important contributions to the payload and the ground segment from Austria, Belgium, France, Germany, Hungary, Italy, Portugal, Spain, Sweden, and the United Kingdom. The CHEOPS Consortium would like to gratefully acknowledge the support received by all the agencies, offices, universities, and industries involved. Their flexibility and willingness to explore new approaches were essential to the success of this mission. CHEOPS data analysed in this article will be made available in the CHEOPS mission archive (\url{https://cheops.unige.ch/archive_browser/}). 
K.W.F.L. was supported by Deutsche Forschungsgemeinschaft grants RA714/14-1 within the DFG Schwerpunkt SPP 1992, Exploring the Diversity of Extrasolar Planets. 
ABr was supported by the SNSA. 
ACMC acknowledges support from the FCT, Portugal, through the CFisUC projects UIDB/04564/2020 and UIDP/04564/2020, with DOI identifiers 10.54499/UIDB/04564/2020 and 10.54499/UIDP/04564/2020, respectively. 
A.C., A.D., B.E., K.G., and J.K. acknowledge their role as ESA-appointed CHEOPS Science Team Members. 
DG gratefully acknowledges financial support from the CRT foundation under Grant No. 2018.2323 “Gaseousor rocky? Unveiling the nature of small worlds”. 
This work has been carried out within the framework of the NCCR PlanetS supported by the Swiss National Science Foundation under grants 51NF40\_182901 and 51NF40\_205606. 
S.G.S. acknowledge support from FCT through FCT contract nr. CEECIND/00826/2018 and POPH/FSE (EC). 
The Portuguese team thanks the Portuguese Space Agency for the provision of financial support in the framework of the PRODEX Programme of the European Space Agency (ESA) under contract number 4000142255. 
TWi acknowledges support from the UKSA and the University of Warwick. 
YAl acknowledges support from the Swiss National Science Foundation (SNSF) under grant 200020\_192038. 
RAl, DBa, EPa, IRi, and EVi acknowledge financial support from the Agencia Estatal de Investigación of the Ministerio de Ciencia e Innovación MCIN/AEI/10.13039/501100011033 and the ERDF “A way of making Europe” through projects  PID2021-125627OB-C31, PID2021-125627OB-C32, PID2021-127289NB-I00, PID2023-150468NB-I00 and PID2023-149439NB-C41. 
SCCB acknowledges the support from Fundação para a Ciência e Tecnologia (FCT) in the form of work contract through the Scientific Employment Incentive program with reference 2023.06687.CEECIND and DOI 10.54499/2023.06687.CEECIND/CP2839/CT0002. 
LBo, VNa, IPa, GPi, RRa, and GSc acknowledge support from CHEOPS ASI-INAF agreement n. 2019-29-HH.0. 
CBr and ASi acknowledge support from the Swiss Space Office through the ESA PRODEX program. 
ACC acknowledges support from STFC consolidated grant number ST/V000861/1, and UKSA grant number ST/X002217/1. 
P.E.C. is funded by the Austrian Science Fund (FWF) Erwin Schroedinger Fellowship, program J4595-N. 
This project was supported by the CNES. 
A.De. 
This work was supported by FCT - Funda\c{c}\~{a}o para a Ci\^{e}ncia e a Tecnologia through national funds and by FEDER through COMPETE2020 through the research grants UIDB/04434/2020, UIDP/04434/2020, 2022.06962.PTDC. 
O.D.S.D. is supported in the form of work contract (DL 57/2016/CP1364/CT0004) funded by national funds through FCT. 
B.-O. D. acknowledges support from the Swiss State Secretariat for Education, Research and Innovation (SERI) under contract number MB22.00046. 
ADe, BEd, KGa, and JKo acknowledge their role as ESA-appointed CHEOPS Science Team Members. 
This project has received funding from the Swiss National Science Foundation for project 200021\_200726. It has also been carried out within the framework of the National Centre of Competence in Research PlanetS supported by the Swiss National Science Foundation under grant 51NF40\_205606. The authors acknowledge the financial support of the SNSF. 
MF and CMP gratefully acknowledge the support of the Swedish National Space Agency (DNR 65/19, 174/18). 
M.G. is an F.R.S.-FNRS Senior Research Associate. 
MNG is the ESA CHEOPS Project Scientist and Mission Representative. BMM is the ESA CHEOPS Project Scientist. KGI was the ESA CHEOPS Project Scientist until the end of December 2022 and Mission Representative until the end of January 2023. All of them are/were responsible for the Guest Observers (GO) Programme. None of them relay/relayed proprietary information between the GO and Guaranteed Time Observation (GTO) Programmes, nor do/did they decide on the definition and target selection of the GTO Programme. 
CHe acknowledges financial support from the Österreichische Akademie 1158 der Wissenschaften and from the European Union H2020-MSCA-ITN-2019 1159 under Grant Agreement no. 860470 (CHAMELEON). Calculations were performed using supercomputer resources provided by the Vienna Scientific Cluster (VSC). 
This work was granted access to the HPC resources of MesoPSL financed by the Region Ile de France and the project Equip@Meso (reference ANR-10-EQPX-29-01) of the programme Investissements d'Avenir supervised by the Agence Nationale pour la Recherche. 
This work has been carried out within the framework of the NCCR PlanetS supported by the Swiss National Science Foundation under grants 51NF40\_182901 and 51NF40\_205606. AL and JK acknowledge support of the Swiss National Science Foundation under grant number  TMSGI2\_211697. 
ML acknowledges support of the Swiss National Science Foundation under grant number PCEFP2\_194576. 
PM acknowledges support from STFC research grant number ST/R000638/1. 
This work was also partially supported by a grant from the Simons Foundation (PI Queloz, grant number 327127). 
NCSa acknowledges funding by the European Union (ERC, FIERCE, 101052347). Views and opinions expressed are however those of the author(s) only and do not necessarily reflect those of the European Union or the European Research Council. Neither the European Union nor the granting authority can be held responsible for them. 
GyMSz acknowledges the support of the Hungarian National Research, Development and Innovation Office (NKFIH) grant K-125015, a a PRODEX Experiment Agreement No. 4000137122, the Lend\"ulet LP2018-7/2021 grant of the Hungarian Academy of Science and the support of the city of Szombathely. 
V.V.G. is an F.R.S-FNRS Research Associate. 
JV acknowledges support from the Swiss National Science Foundation (SNSF) under grant PZ00P2\_208945. 
NAW acknowledges UKSA grant ST/R004838/1. 
YNEE acknowledges support from a Science and Technology Facilities Council (STFC) studentship, grant number ST/Y509693/1.

The Belgian participation to CHEOPS has been supported by the Belgian Federal Science Policy Office (BELSPO) in the framework of the PRODEX Program, and by the University of Liège through an ARC grant for Concerted Research Actions financed by the Wallonia-Brussels Federation.

\end{acknowledgements}

\bibliographystyle{aa} 
\bibliography{biblio} 

\clearpage\newpage
\begin{appendix}

\section{Additional figures}


\begin{figure*}[ht!]
    \centering
    \subfigure[Visit 1]{\includegraphics[width=0.49\textwidth]{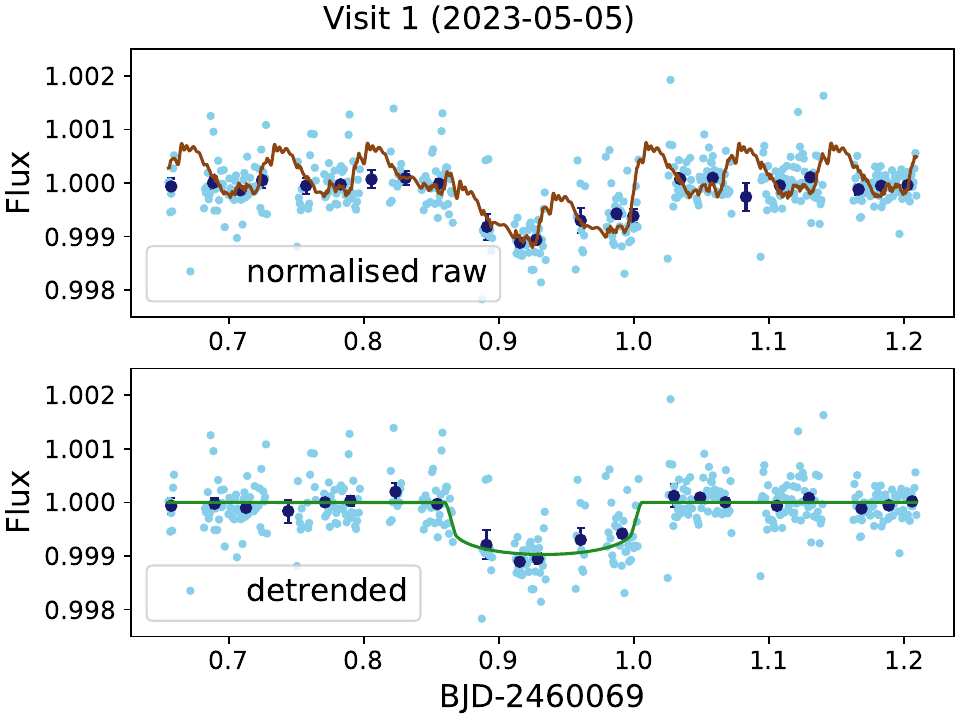}} 
    \subfigure[Visit 2]{\includegraphics[width=0.49\textwidth]{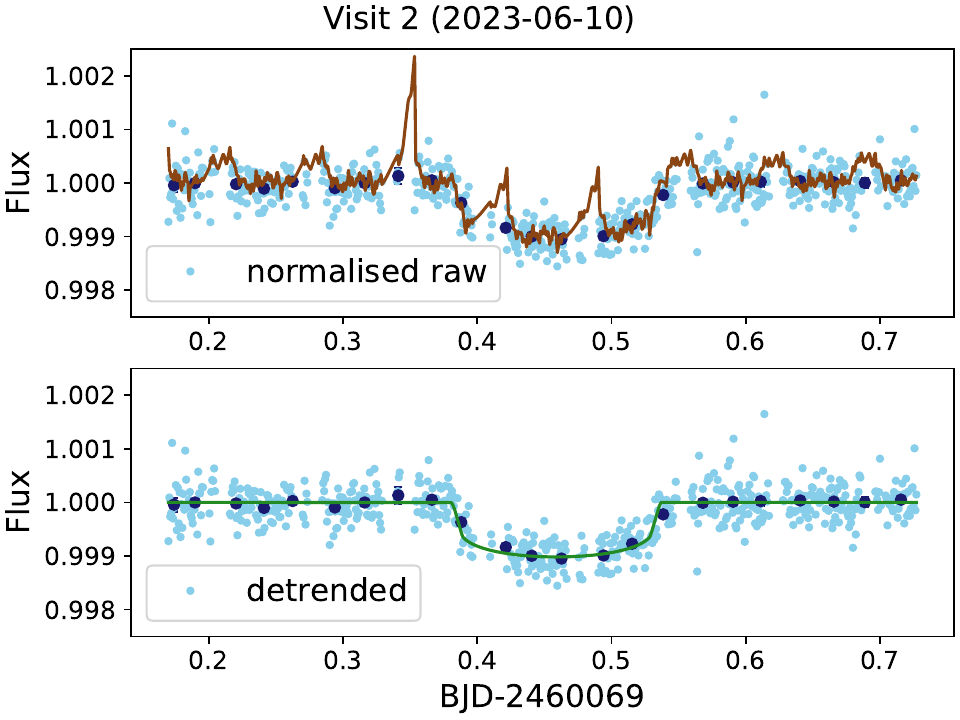}} \\
    \subfigure[Visit 3]{\includegraphics[width=0.49\textwidth]{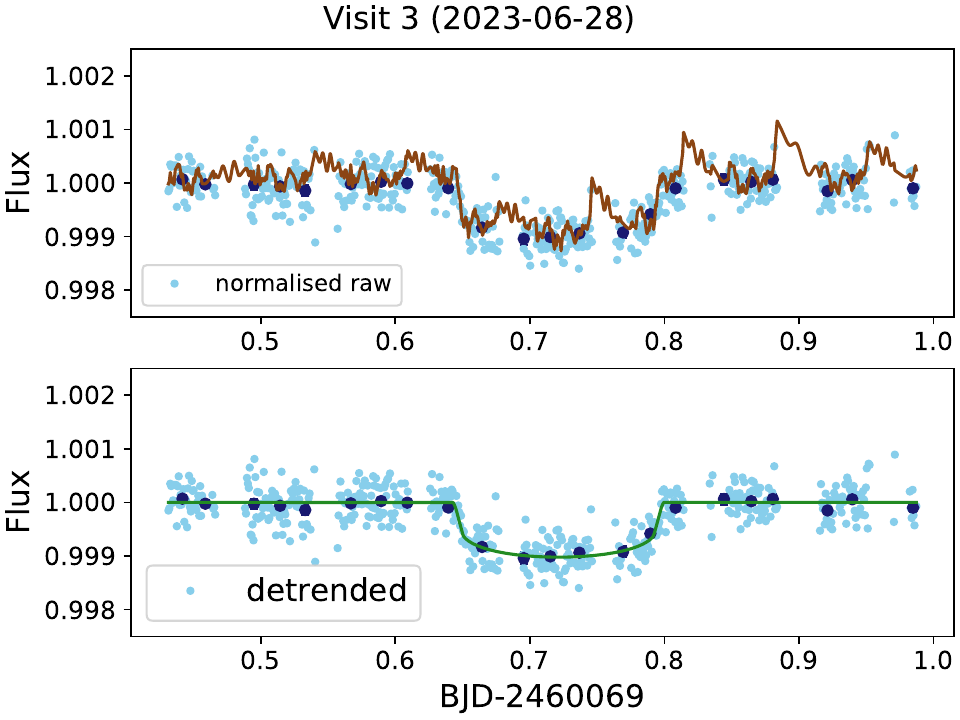}} 
    \subfigure[Visit 4]{\includegraphics[width=0.49\textwidth]{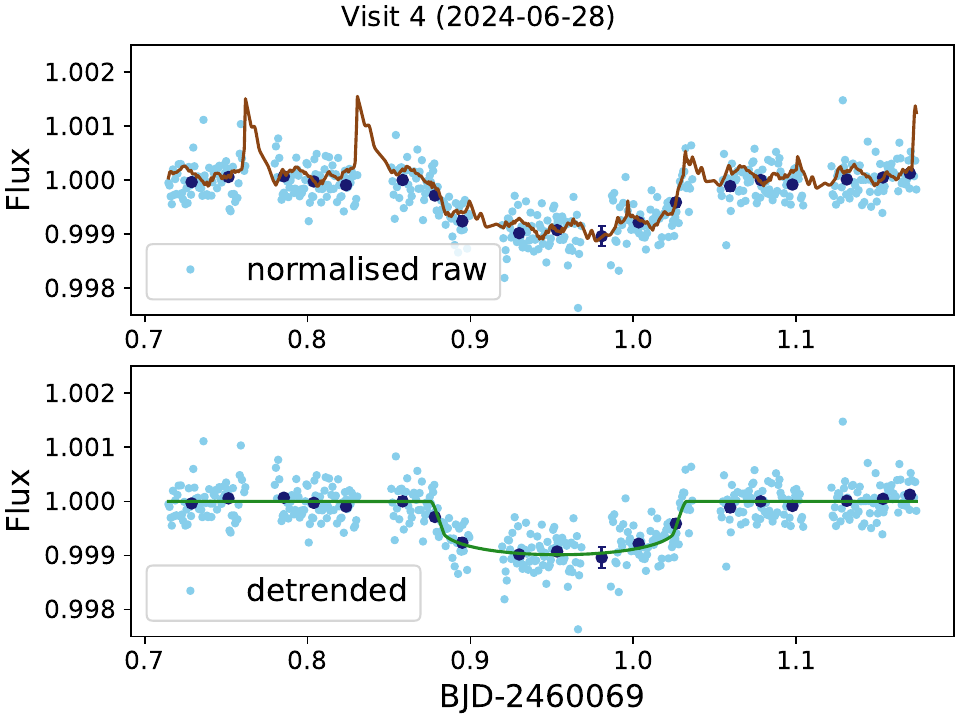}} \\
    \subfigure[Visit 5]{\includegraphics[width=0.49\textwidth]{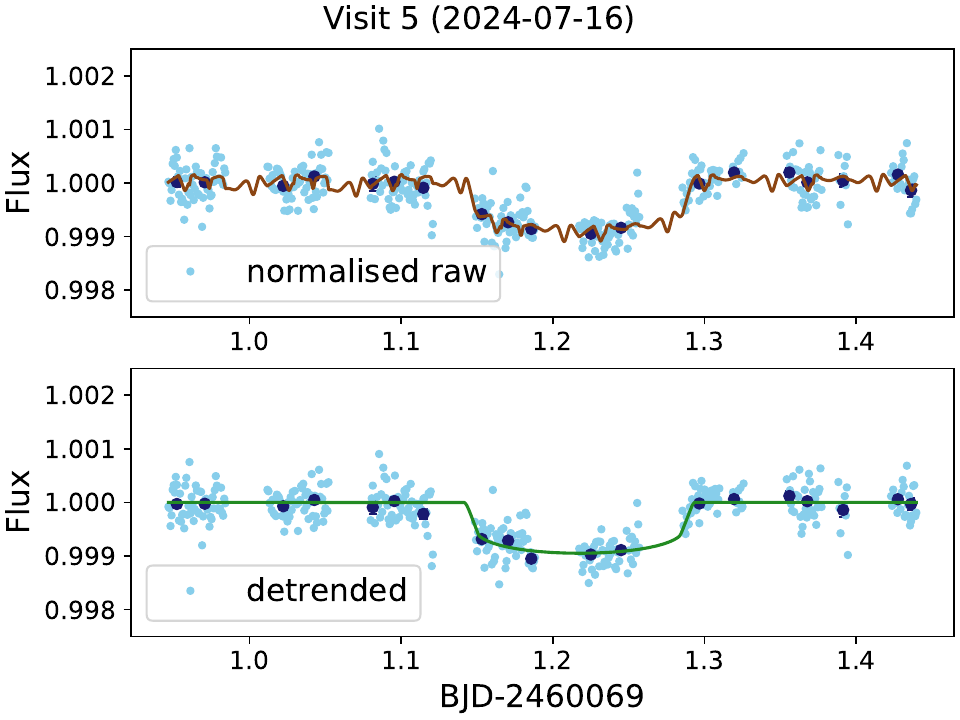}}
    \caption{CHEOPS photometry of TOI-2141 analysed using \texttt{pycheops} (see text). The transits of planet c were observed in five separate visits. Top panels show the normalised raw CHEOPS light curve (sky blue), the binned raw light curve (dark blue), and the best fitted GP+transit model (brown line). Bottom panel shows the detrended light curves (sky blue), the binned, detrended light curve and the best-fitted transit model (green).}
    \label{fig:cheops_raw}
\end{figure*}


\begin{figure*}[ht!]
    \centering
    \includegraphics[width=\hsize]{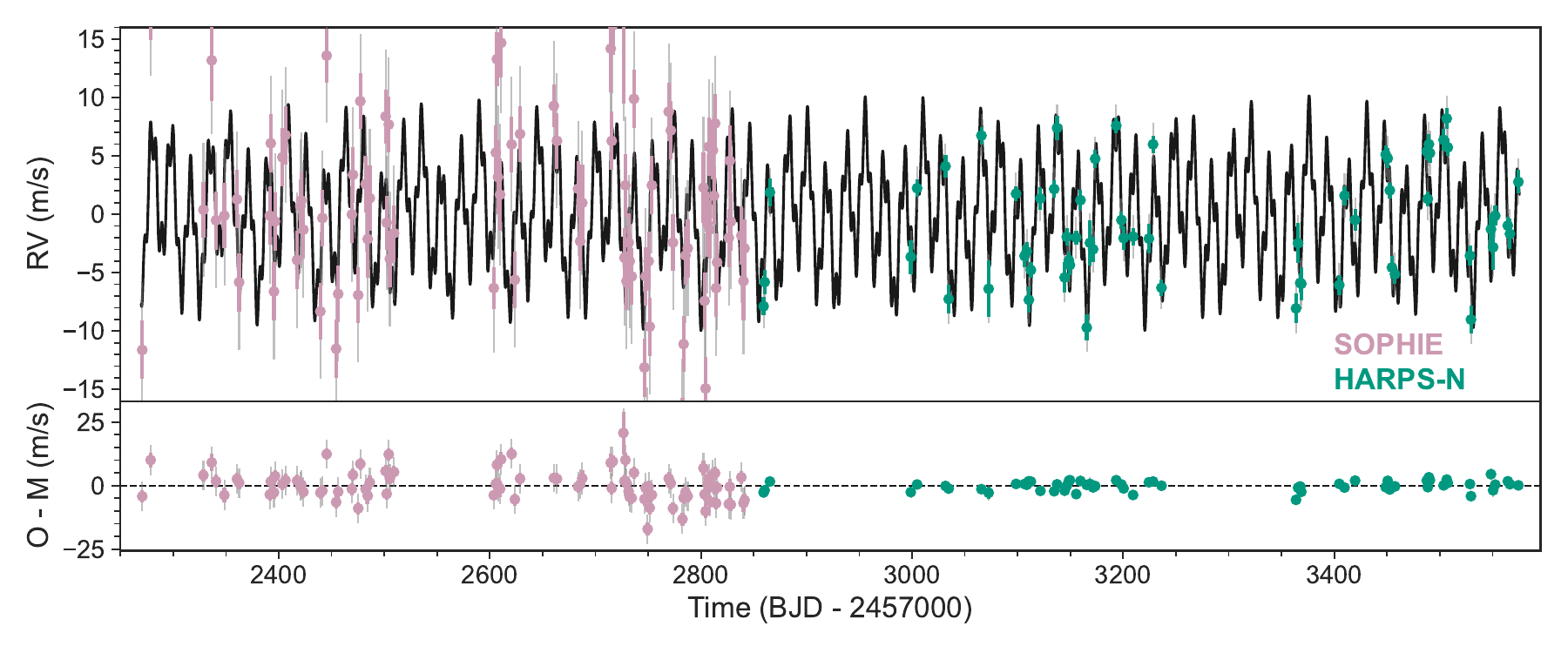}
    \caption{RV time series from SOPHIE (pink) and HARPS-N (green). The error bars are broken into the internal measurement error and the added jitter in gray. The black line shows the best-fit model from Sect.~\ref{subsec:joint_fit}.}
    \label{fig:rvs}
\end{figure*}


\begin{figure}
    \centering
    \includegraphics[width=\hsize]{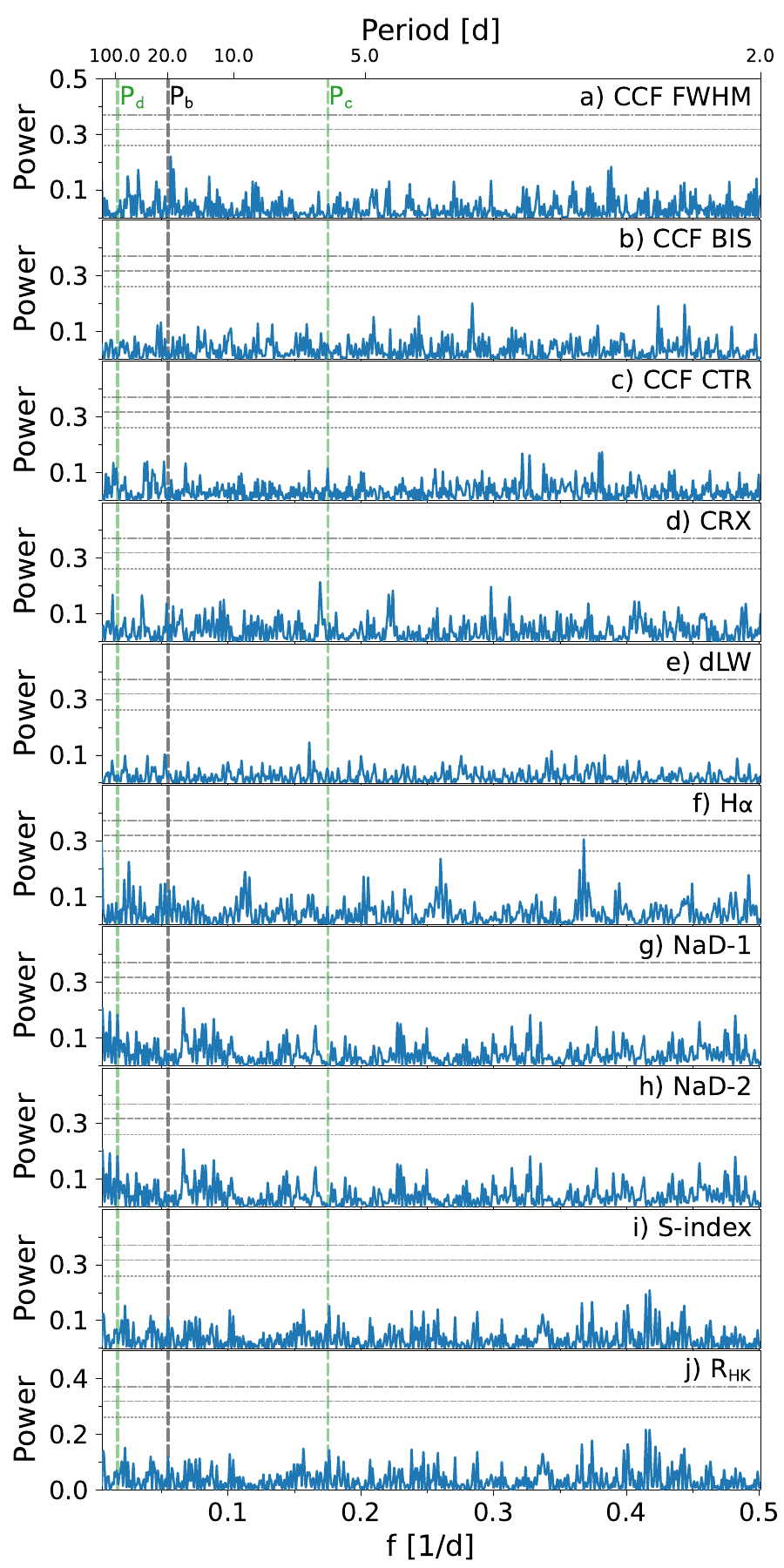}
    \caption{GLS periodograms of the spectral activity indicators from HARPS-N data. For each panel, the horizontal lines show the theoretical 10\% (short-dashed line), 1\% (long-dashed line), and 0.1\% (dot-dashed line) false alarm probability levels. The vertical dashed lines mark the orbital frequencies of the transiting planet ($f_\mathrm{b}=0.0547\,{\rm d}^{-1}$) and of the signals detected in the RVs at 5.46 and 60\,d. \emph{Panels a--c}: Cross-correlation full width half maximum (FWHM), bisector (BIS), and contrast (CTR) computed with YABI. \emph{Panels d--h}: Chromatic index (CRX), differential line width (dLW), H$\alpha$, and Na doublet computed with \texttt{serval}. \emph{Panels i--j}: S-index and $R_{HK}^\prime$ index computed by YABI.}
    \label{fig:gls_activity}
\end{figure}


\begin{figure*}
    \centering
    \includegraphics[width=\hsize]{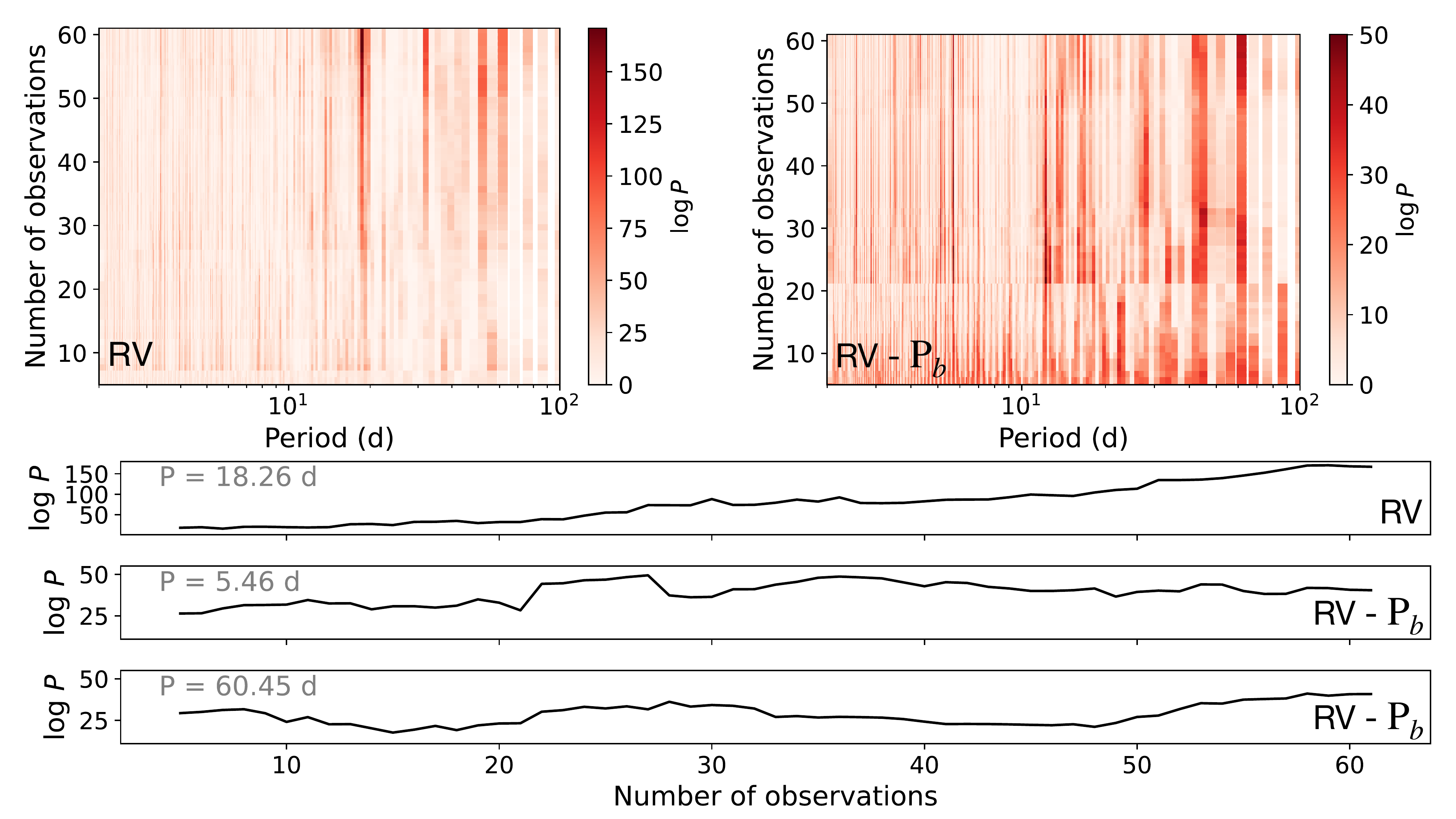}
    \caption{Stacked Bayesian GLS periodogram analysis of the HARPS-N dataset. \textit{Top}: Periodogram of the original HARPS-N RV dataset (left) and the HARPS-N RV dataset after modeling of the transiting planet TOI-2141~b (right). The amount of observations is plotted against period, with the color scale indicating the logarithm of the the power or probability of a coherent long-lived signal, where darker is more likely. \textit{Bottom}: Power as a function of the number of observations for the three RV signals of interest in the system: the transiting planet at 18.26\,d (top), the inner non-transiting planet at 5.46\,d (middle), and the outer non-transiting planet at 60.45\,d (bottom). The first panel is computed from the original RV dataset, the next two from the residuals of fitting the transiting planet. The monotonic increase of power as afunction of number of observations is typical of bonafide planets and not of astrophysical or instrumental origin \citep{SBGLS}. } 
    \label{fig:sbgls}
\end{figure*}


\begin{figure*}[ht!]
    \centering
    \includegraphics[width=0.49\hsize]{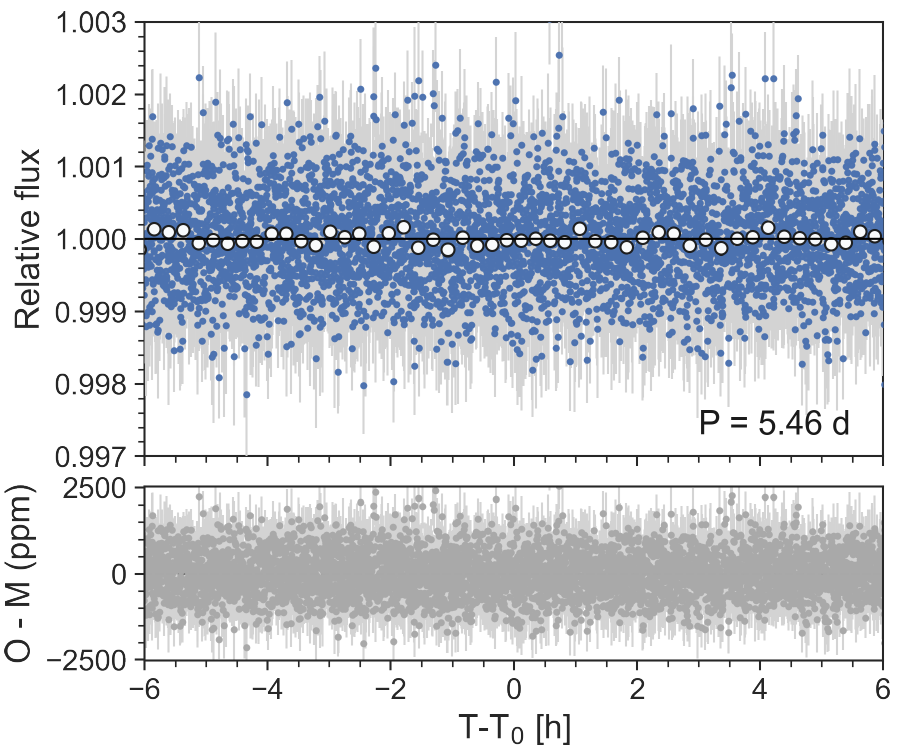}
    \includegraphics[width=0.49\hsize]{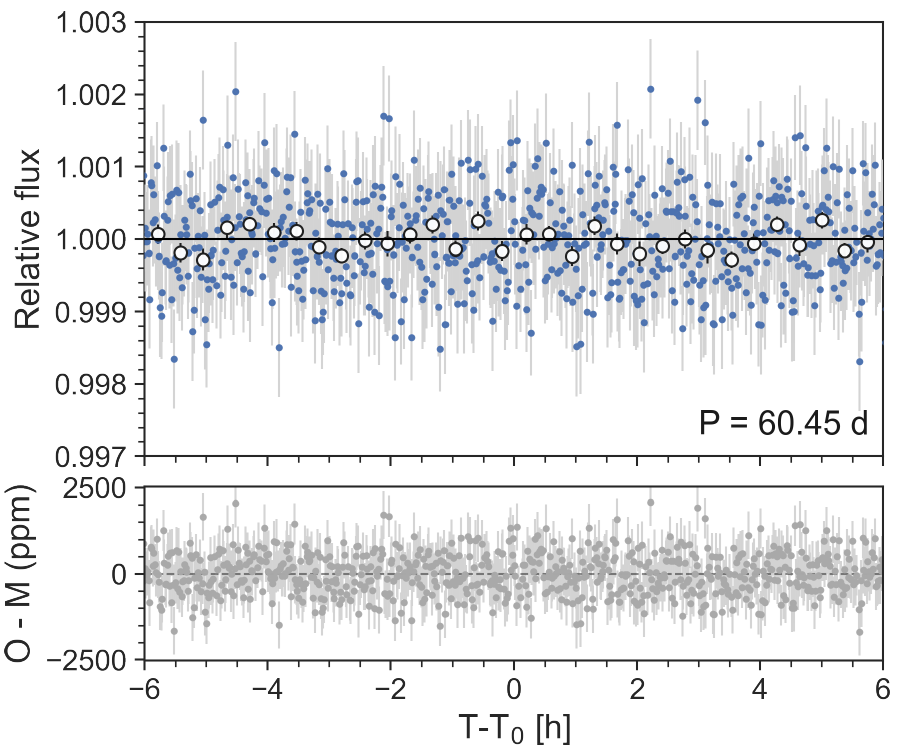}
    \caption{Phase-folded TESS photometry to the period of planet c (left) and d (right). White circles represent the binned photometry for readability. None of the planets appears to be transiting during the time of our observations.}
    \label{fig:nontransiting}
\end{figure*}

\end{appendix}

\end{document}